\documentclass[fleqn,usenatbib]{mnras}

\usepackage{newtxtext,newtxmath}

\usepackage[T1]{fontenc}

\DeclareRobustCommand{\VAN}[3]{#2}
\let\VANthebibliography\thebibliography
\def\thebibliography{\DeclareRobustCommand{\VAN}[3]{##3}\VANthebibliography}

\usepackage{graphicx}	
\usepackage{amsmath}	
\usepackage{amssymb}	

\DeclareUnicodeCharacter{2212}{-}

\title[Serendipitous M dwarf radio flaring]{Serendipitous discovery of radio flaring behaviour from a nearby M dwarf with MeerKAT}

\author[A. Andersson et al.]{
Alex Andersson,$^{1}$\thanks{E-mail: alexander.andersson@physics.ox.ac.uk}
Rob P. Fender,$^{1,2}$ 
Chris J. Lintott,$^{1}$  
David R.A. Williams,$^{3}$ 
Laura N. Driessen,$^{3,4}$ 
\newauthor{
Patrick A. Woudt,$^{2}$  
Alexander J. van der Horst,$^{5,6}$ 
David A.H. Buckley,$^{7}$ 
Sara E. Motta,$^{8}$ 
Lauren Rhodes$^{1,9}$ 
}
\newauthor{
Nora L. Eisner,$^{1}$ 
Rachel A. Osten,$^{10,11}$ 
Paul Vreeswijk,$^{12}$ 
Steven Bloemen$^{12}$
and Paul J. Groot$^{2,7,12}$  
}
\\
$^{1}$Department of Physics, Astrophysics, University of Oxford, Denys Wilkinson Building, Keble Road, Oxford OX1 3RH, UK\\
$^{2}$Department of Astronomy, University of Cape Town, Private Bag X3, Rondebosch 7701, South Africa\\
$^{3}$Jodrell Bank Centre for Astrophysics, Department of Physics and Astronomy, The University of Manchester, Manchester, M13 9PL, UK \\
$^{4}$CSIRO, Space and Astronomy, PO Box 1130, Bentley, WA 6102, Australia\\
$^{5}$Department of Physics, The George Washington University, 725 21st Street NW, Washington, DC 20052, USA\\
$^{6}$Astronomy, Physics and Statistics Institute of Sciences (APSIS), 725 21st Street NW, Washington, DC 20052, USA\\
$^{7}$South African Astronomical Observatory, PO Box 9, Observatory 7935, South Africa\\
$^{8}$Istituto Nazionale di Astrofisica, Osservatorio Astronomico di Brera, via E. Bianchi 46, 23807 Merate (LC), Italy\\
$^{9}$Max-Planck-Institut für Radioastronomie, Auf dem Hügel 69, 53121 Bonn, Germany\\
$^{10}$Space Telescope Science Institute, 3700 San Martin Drive, Baltimore, MD 21218, USA\\
$^{11}$Center for Astrophysical Sciences, Johns Hopkins University, Baltimore, MD, USA\\
$^{12}$Department of Astrophysics/IMAPP, Radboud University, P.O. 9010, 6500 GL, Nĳmegen, The Netherlands
}

\date{Accepted XXX. Received YYY; in original form ZZZ}

\pubyear{2021}

\begin{document}
\label{firstpage}
\pagerange{\pageref{firstpage}--\pageref{lastpage}}
\maketitle

\begin{abstract}
We report on the detection of MKT J174641.0$-$321404, a new radio transient found in untargeted searches of wide-field MeerKAT radio images centred on the black hole X-ray binary H1743$-$322. MKT J174641.0$-$321404 is highly variable at 1.3 GHz and was detected three times during 11 observations of the field in late 2018, reaching a maximum flux density of 590 $\pm$ 60 $\mu$Jy. We associate this radio transient with a high proper motion, M dwarf star SCR~1746$-$3214 12 pc away from the Sun. Multiwavelength observations of this M dwarf indicate flaring activity across the electromagnetic spectrum, consistent with emission expected from dMe stars, and providing upper limits on quiescent brightness in both the radio and X-ray regimes. \textit{TESS} photometry reveals a rotational period for SCR~1746$-$3214 of $0.2292 \pm 0.0025$ days, which at its estimated radius makes the star a rapid rotator, comparable to other low mass systems. Dedicated spectroscopic follow up confirms the star as a mid-late spectral M dwarf with clear magnetic activity indicated by strong H$\alpha$ emission. This transient's serendipitous discovery by MeerKAT, along with multiwavelength characterisation, make it a prime demonstration of both the capabilities of the current generation of radio interferometers and the value of simultaneous observations by optical facilities such as MeerLICHT. Our results build upon the literature of of M dwarfs’ flaring behaviour, particularly relevant to the habitability of their planetary systems.

\end{abstract}





\section{Introduction}
\label{sec:intro}
Radio telescopes have, until recently, been primarily of use for variable and transient searches as follow-up instruments, typically informed by detections at higher energies such as optically transient supernovae, gamma-ray bursts (GRBs) and X-ray binaries (XRBs). However, with the advent of the current generation of wide field radio interferometers, including Square Kilometre Array (SKA) pathfinder instruments like MeerKAT \citep{Camilo2018}, the Australian SKA Pathfinder \citep[ASKAP;][]{Johnston2007}, the Murchison Widefield Array \citep[MWA;][] {Tingay2012} and the LOw Frequency ARray \citep[LOFAR;][]{VanHaarlem2013}, the astronomical community can now sample the radio sky with improved sensitivity and, crucially, with larger fields of view (FoV).

To date, the number of serendipitously detected, image plane radio transients remains low. This is in contrast to the well established coherent population of variable radio sources such as pulsars and fast radio bursts (see \citealp{Pietka2015} and \citealp{Fender2011} for a discussion on coherent and incoherent transients) that are discovered regularly in untargeted searches. Ongoing image plane searches for radio transients include the ASKAP survey for Variables and Slow Trasients \citep[VAST;][]{Murphy2013} and the Amsterdam--ASTRON Radio Transients Facility and Analysis Centre \citep[AARTFAAC;][]{Prasad2016}, along with large field surveys such as the Karl G. Jansky Very Large Array's (VLA) Sky Survey \citep[VLASS;][]{Lacy2019}. A number of transients have been discovered at frequencies ranging from tens of MHz to a few GHz for which there have yet to be associated multiwavelength counterparts or definite progenitor systems \citep[e.g.][]{Hyman2005, Bower2007, Ofek2011, Jaeger2012, Stewart2016, Murphy2017, Varghese2019, 2021ApJ...920...45W}. Large surveys have shown that, in the radio band, many variable and transient phenomena can be attributed to scintillation or intrinsic AGN variation (e.g. \citealp{Levinson2002}; \citealp{Bannister2011}; \citealp{Mooley2016}; \citealp{Bhandari2018}; \citealp{Radcliffe2019}; \citealp{2022MNRAS.tmp..746D}). For example, \cite{Sarbadhicary2020} utilise the VLA's COSMOS HI Legacy Survey to create a Variable and Explosive Radio Dynamic Evolution Survey (CHILES VERDES), which makes use of the plethora of multiwavelength data of the COSMOS field and 5.5 years worth of data reaching RMS sensitivities of $\sim 10   \mu$Jy beam$^{-1}$ per epoch. Their results reach low flux density limits and find 58 AGN-type sources using the moderate FoV of radius 22.5'. 
From this literature, only $\sim1-5$ per cent of the radio sky seems to be variable, with the number of confirmed galactic sources limited primarily to follow-up or targeted observations of known variables. \cite{Mooley2016} associate two of their transients with known types of active star whilst \cite{Driessen2020} describe the first commensally discovered image plane transient from MeerKAT, MKT J170456.2$-$482100, similarly associating it with a known stellar system. One of the common conclusions arrived at by much of the radio variability literature is that, in order to systematically find radio transients, high sensitivity, regular cadence and large FoVs are required and that surveys lacking any of these three pillars can limit detection capabilities. 


ThunderKAT\footnote{The HUNt for Dynamic and Explosive Radio transients with MeerKAT} \citep{Fender2016} is the MeerKAT Large Survey Project (LSP) dedicated to image plane radio transients. This project uses MeerKAT's 64 dishes and FoV > 1 square degree to take regular cadence observations of reported transients such as XRBs, cataclysmic variables and GRBs, whilst also operating commensal searches of internal and cross-LSP data. ThunderKAT also makes use of the robotic optical facility MeerLICHT \citep[65\,cm primary mirror;][]{Bloemen2016} which was built and is operated to support transient work with MeerKAT. MeerLICHT observes simultaneously with  night-time MeerKAT observations, allowing for better characterisation of the multiwavelength transient sky. Both dedicated and commensal work from the ThunderKAT team is bearing fruit and herein we follow in the footsteps of \cite{Driessen2020}  by reporting on the discovery of MKT J174641.0$-$321404, the second serendipitous, image plane radio transient found by MeerKAT, and associate it with a known, nearby red dwarf.

Radio transients have been seen originating from some late-type dwarf stars, typically as a result of magnetic activity and reconnection. These systems are sometimes referred to as ‘dMe' stars due to the near ubiquitous presence of hydrogen emission lines, indicative of chromospheric heating (\citealp{Cram1979}; \citealp{Cram1987}); for reviews of stellar radio astronomy across the Hertzsprung--Russell diagram and on dwarf flares see \cite{Gudel2002} and \cite{Osten2007} respectively. Some of the pioneers of radio astronomy observed radio flares from late-type dwarfs, including the prototypical class namesake UV Ceti (\citealp{Lovell1963}; \citealp{Lovell1969}). Stellar radio flares are caused by either incoherent cyclotron or gyrosynchrotron emission of electrons caught in magnetic fields, or coherent bursts attributed to either plasma- or electron cyclotron maser-emission, which probe the local electron density and magnetic field strength respectively. For a detailed review of the emission processes in radio bright stars we refer the reader to \cite{Dulk1985}. Most radio observations of flare stars have focused on a few nearby sources such as UV Ceti, Proxima Centauri, AD Leonis and YZ Canis Minoris (\citealp{Lacy1976}; \citealp{Bastian1990}; \citealp{Slee2003}; \citealp{Villadsen2019}), however \cite{Pritchard2021} utilised the circular polarisation capabilities of ASKAP to detect coherent emission from 23 stars with no previous radio counterparts. This was done with data reaching an RMS sensitivity of 250 $\mu$Jy beam$^{-1}$ across the entire sky south of $+41^{\circ}$ declination, again emphasising the need for wide field observations. ASKAP has also been used to detect elliptically polarized radio pulses from UV Ceti \citep{Zic2019}. In just the past year \cite{Driessen2021} have found radio emission coming from the previously known X-ray flare star EXO 040830−7134.7 - a chromospherically active M-dwarf of spectral type M0V - and \cite{2021NatAs.tmp..196C} detailed the detection of coherent emission from 19 M dwarfs with LOFAR.

In Section \ref{sec:RadObs} we detail the radio observations and source discovery, whilst Section \ref{sec:Multi} reconciles said observations with archival optical and X-ray data. Section \ref{sec:Spect} presents dedicated spectroscopy before the discussion and conclusions of Sections \ref{sec:Disc} and \ref{sec:Conc} respectively.

\section{Radio Observations}
\label{sec:RadObs}

As part of the ThunderKAT monitoring program, XRB H1743$-$322 was observed at weekly cadence for 11 epochs, following reports of new outburst behaviour \citep{Williams2020}.
Weekly monitoring of this field ran from 9th September until the 10th of November 2018, with each observation consisting of 15 minutes on source and an integration time of 8 seconds. The on source observations were preceded and succeeded by 2 minutes on the phase calibration source J1712$-$281 or J1830$-$3602, whilst the  band-pass calibrator PKS J1939$-$6342 was observed for 10 minutes before beginning each observation block. These observations were taken using the L-band (900--1670 MHz) receiver, with a central frequency of 1284 MHz, bandwidth of 856 MHz and 4096 frequency channels. The data were reduced using the standard procedure, from flagging with \texttt{AOFlagger} \citep{Offringa2012} and calibration using \texttt{CASA} \citep{Mcmullin2007} through to imaging with \texttt{WSClean} (\citealp{Offringa2014}; \citealp{Offringa2017}). For full details on the data reduction see \cite{Williams2020}.

\subsection{TraP}
\label{sec:trap} 

The MeerKAT images were run through the LOFAR Transients Pipeline \citep[\texttt{TraP};][]{Swinbank2015} to detect serendipitous varying and transient sources in the large field of view observations. This pipeline does source finding and association across all epochs and constructs a database of sources and their associated light curves. From each source's light curve, variability statistics $\eta$ and $V$ are calculated. The former, defined for $N$ flux density measurements $I \pm \sigma$ of weighted mean $\xi$, as 
\begin{equation}
    \eta=\frac{1}{N-1}\sum_{i=1}^N \frac{(I_i - \xi_{I_N})^2}{\sigma_i^2}
	\label{eq:eta}
\end{equation}

describes their reduced chi-squared value when compared to a stable source i.e. constant flux density sources are expected to have $\eta \approx 1$ whilst variable sources produce values greater than unity. $V$ is sometimes known as the modulation parameter or co-efficient of variability and is the ratio of the sample standard deviation to the mean of its flux measurements. In general, sources with large values for both  $V$ and $\eta$ are likely to be identified as transients or variable. For more information on how \texttt{TraP} processes data and calculates its statistics refer to \cite{Swinbank2015} or other studies using the pipeline (e.g. \citealp{Rowlinson2019}, \citealp{Driessen2020} or \citealp{Sarbadhicary2020}).

\begin{figure}
	\includegraphics[width=\columnwidth]{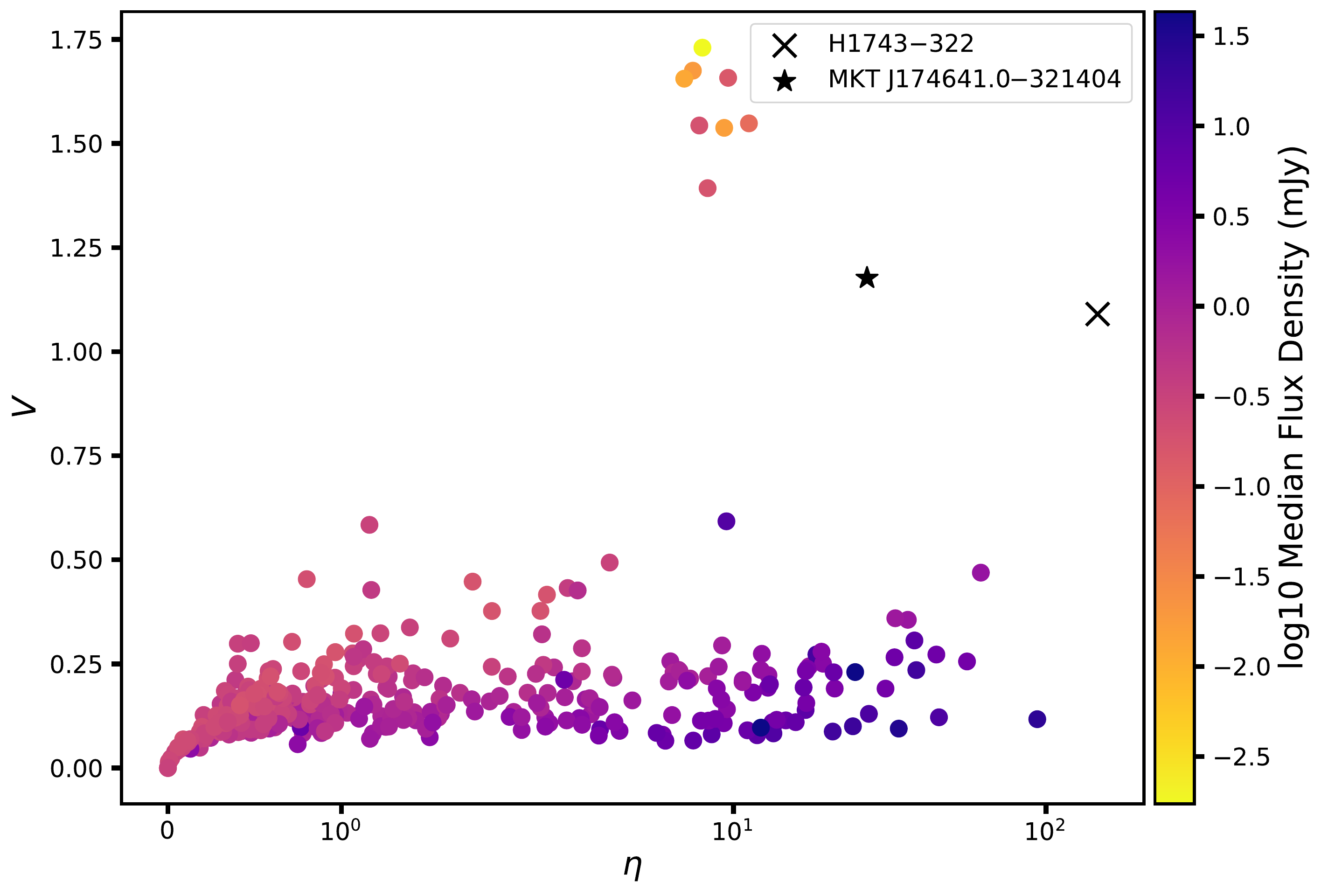}
    \caption{Variability parameters for all sources detected in the H1743$-$322 field during ThunderKAT observations in late 2018. Clear outliers, particularly in $V$ can be seen, including the XRB H1743$-$322 itself, several manually vetted artifacts and an unknown source.}
    \label{fig:etaV}
\end{figure}

The \texttt{TraP} was first run with default parameters, analysing the H1743$-$322 field out to approximately 1.5$\times$ the primary beam radius ($\sim 45'$). During fitting, each source is assumed to be the size of the synthesised beam as here we are only interested in the unresolved point source variability. Using the default 8$\sigma$ detection threshold above the local noise, 377 individual sources were detected across all epochs. This threshold was used as a trade-off between being inundated with many false positives and missing genuine transients. The resultant \texttt{TraP} variability parameter distributions are detailed in Figure~\ref{fig:etaV}, showing outliers to typical $\eta$ and $V$ parameters (i.e. $V \gtrsim 0.75$ and $\eta \gtrsim100$). We note that $\sim50$ per cent of sources lie below $\eta=1$, whilst those above are either bright sources with very little variability ($\eta$ is approximately proportional to signal-to-noise ratio squared), or extended sources whose variation is artificially caused by the shape of the synthesised beam changing between epochs. The grouping of high-$V$ sources were all inspected by eye and found to be imaging or processing artifacts near bright sources. Two highly significant astrophysical sources were identified. One, reassuringly, was the target of this field, the candidate black hole XRB H1743$-$322, and the other was MKT J174641.0−321404, a hitherto unknown radio source located $\sim 5'$ East of the aforementioned XRB.

\begin{figure*}
	\includegraphics[width=.33\textwidth]{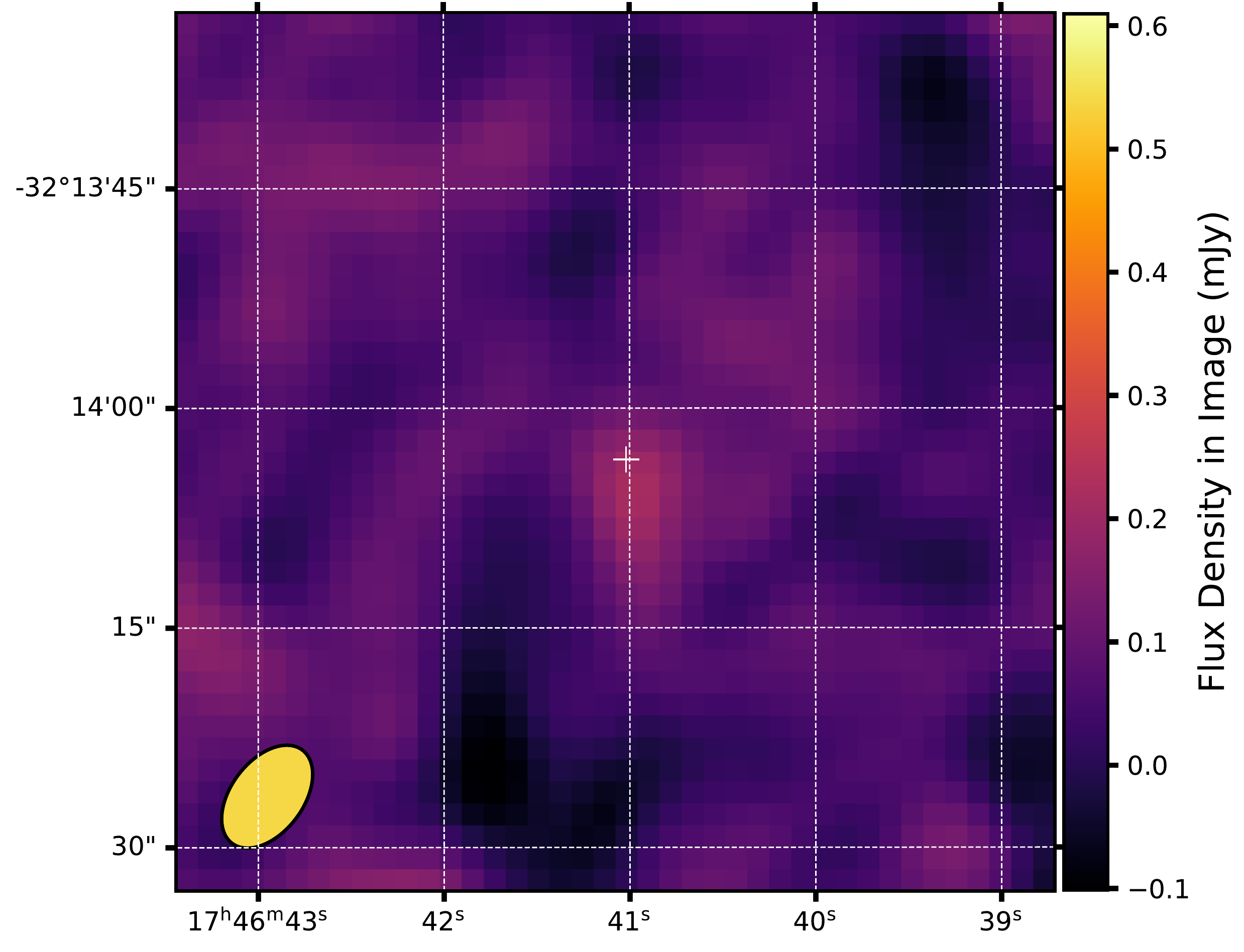}
	\includegraphics[width=.33\textwidth]{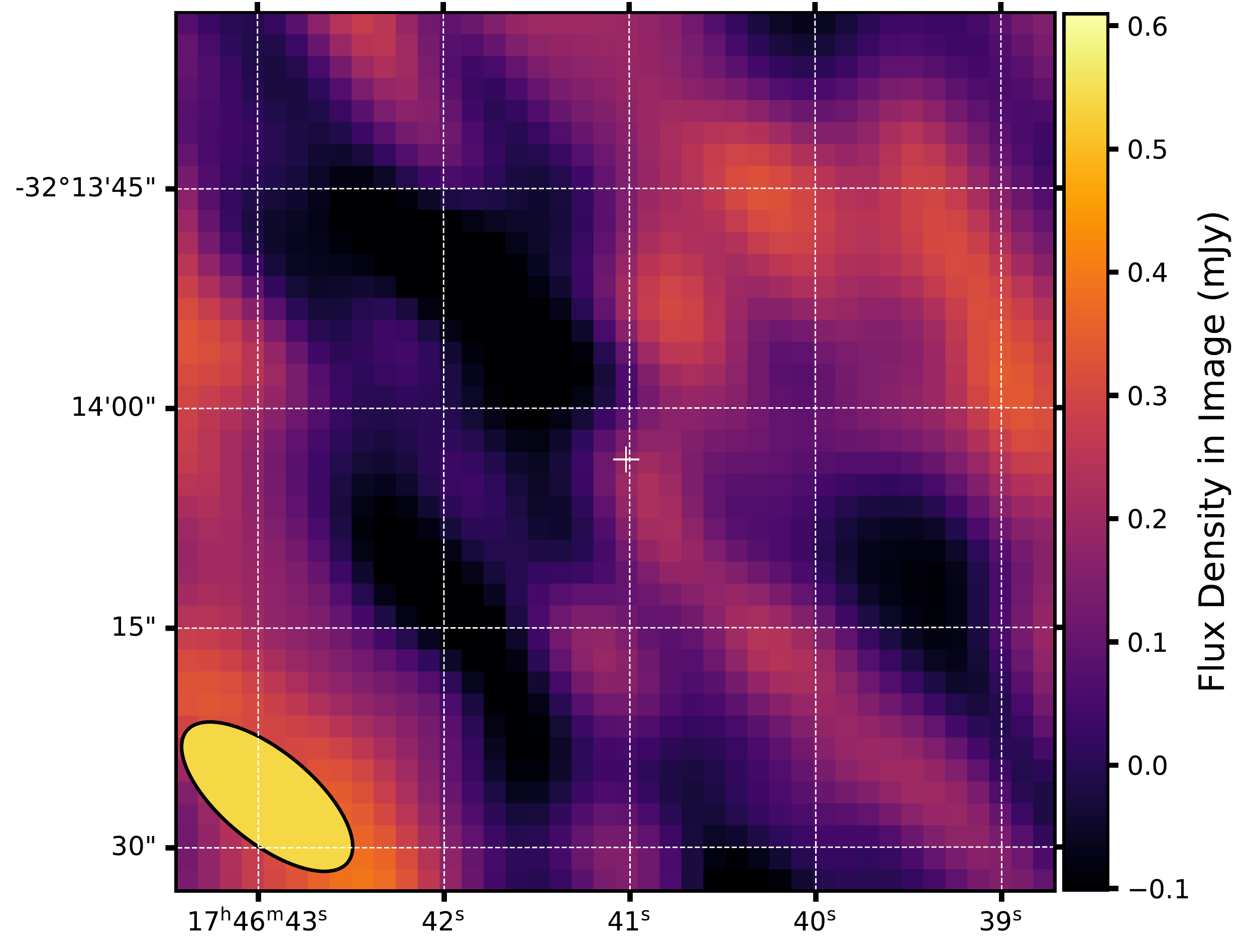}
	\includegraphics[width=.33\textwidth]{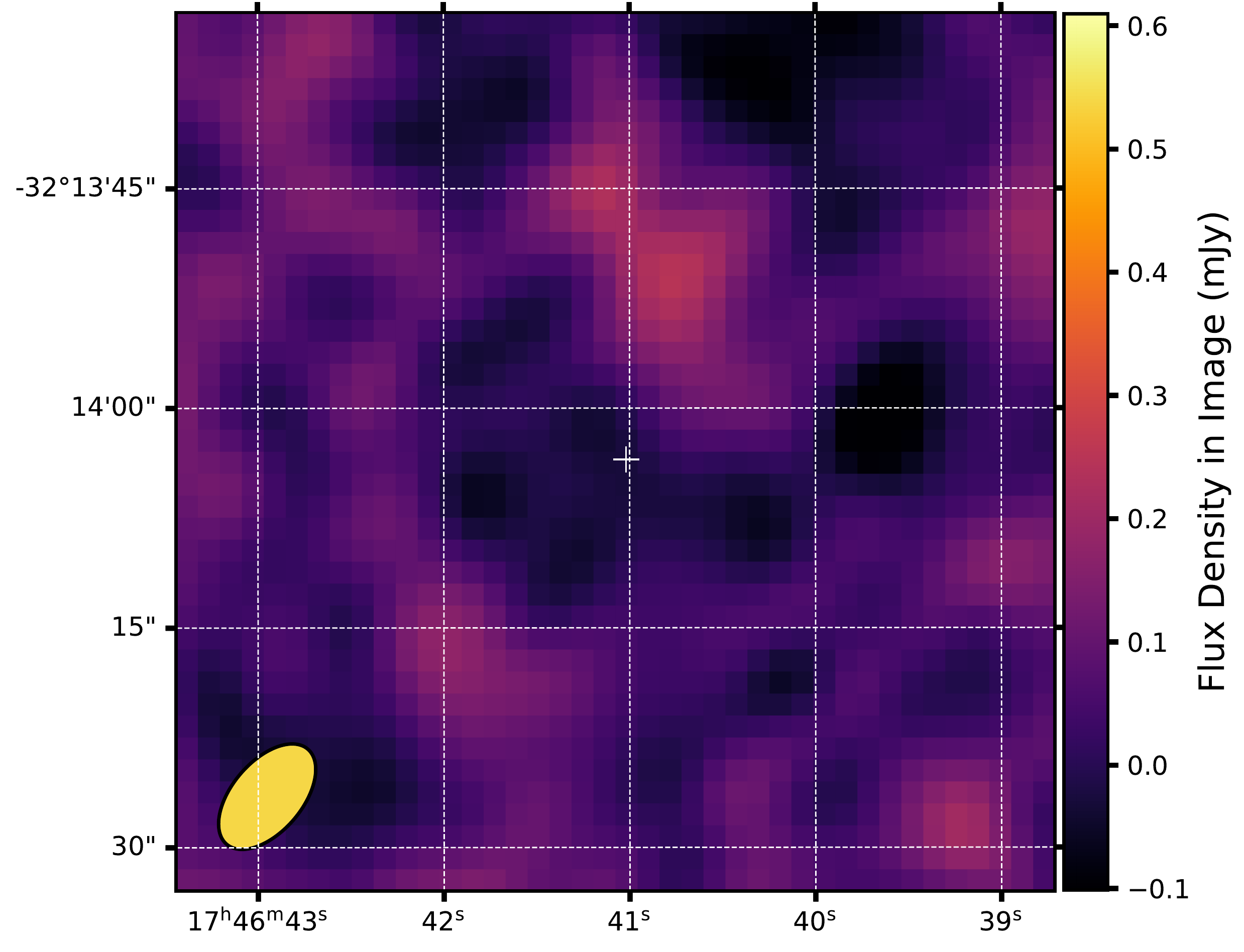}
	\includegraphics[width=.33\textwidth]{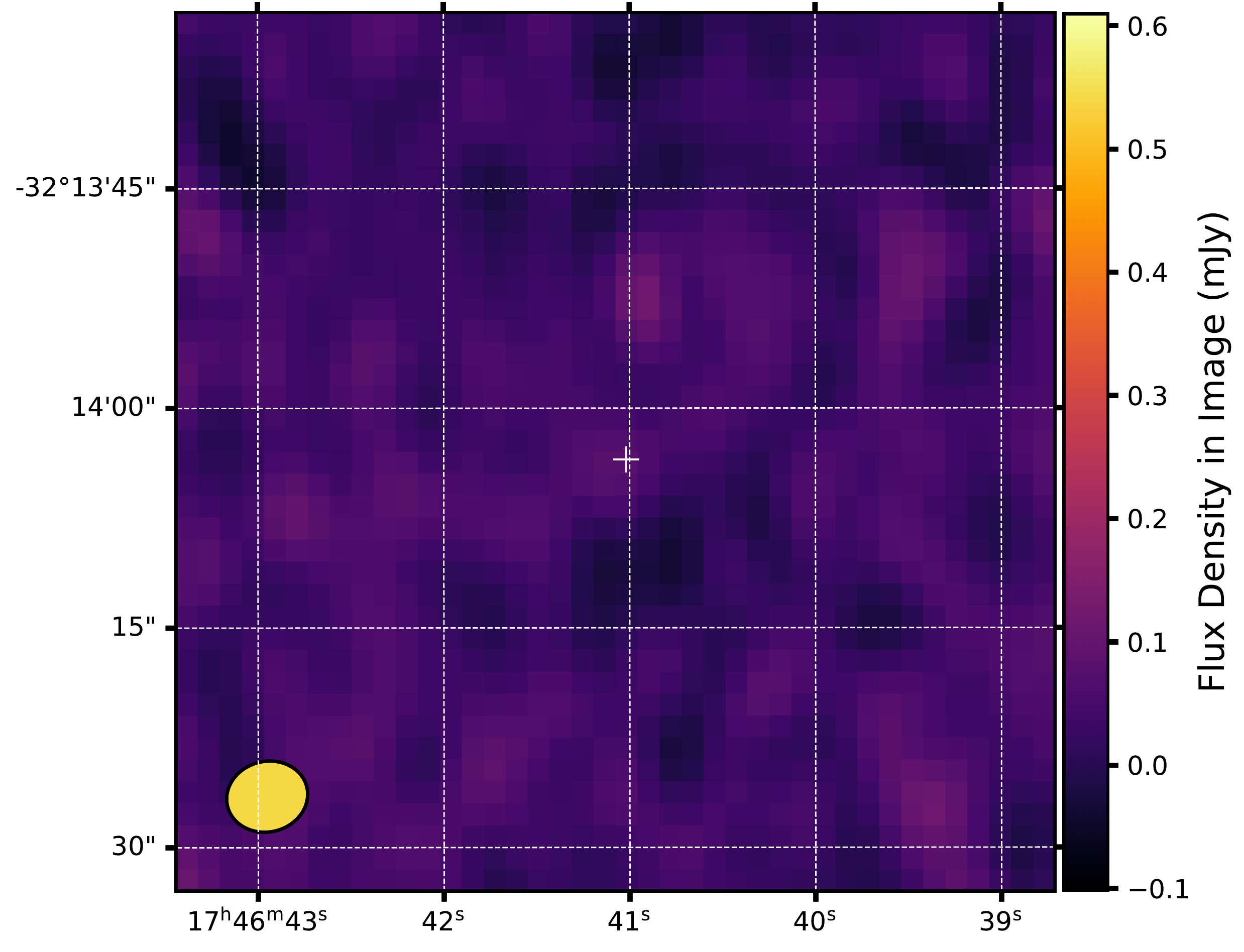}
	\includegraphics[width=.33\textwidth]{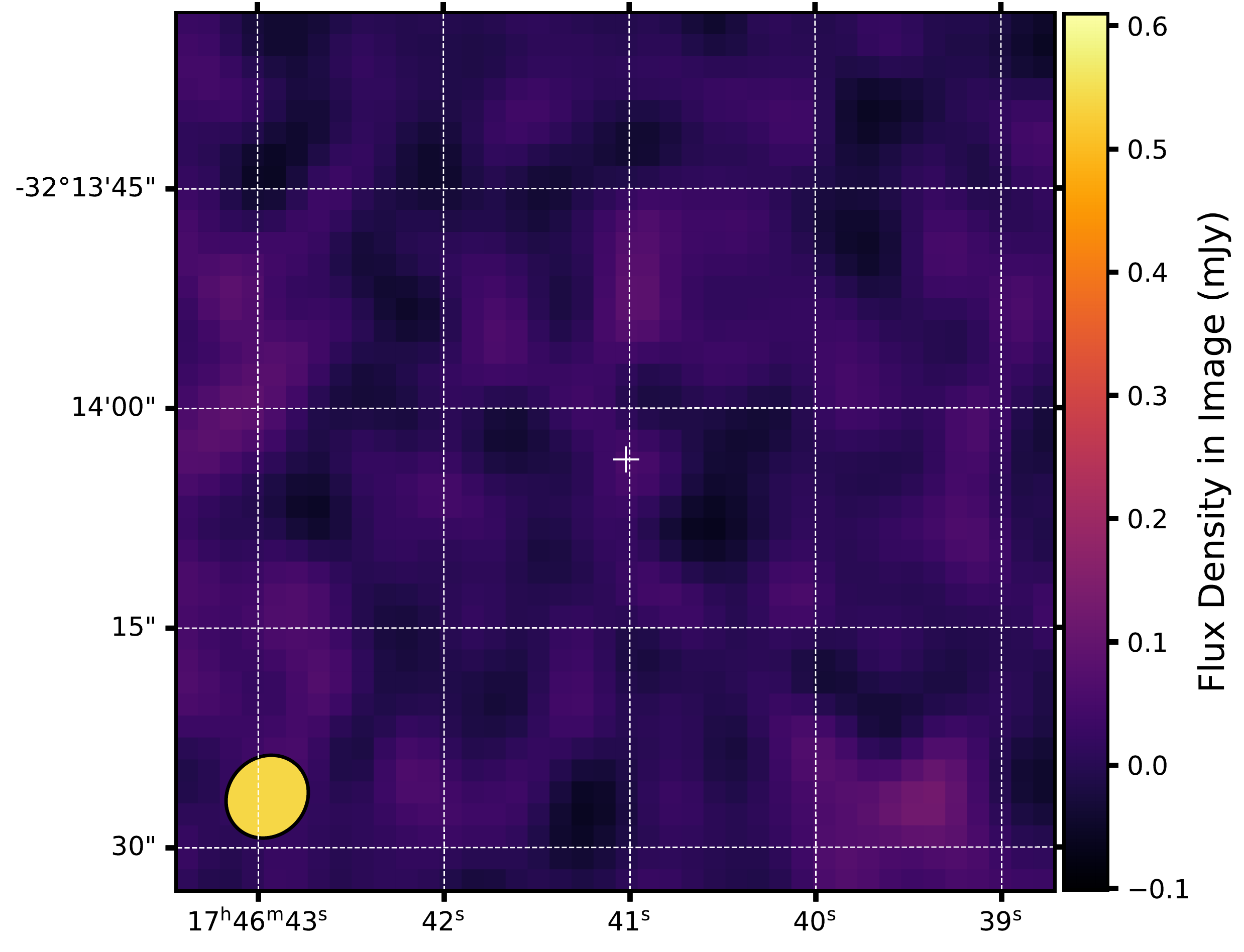}
	\includegraphics[width=.33\textwidth]{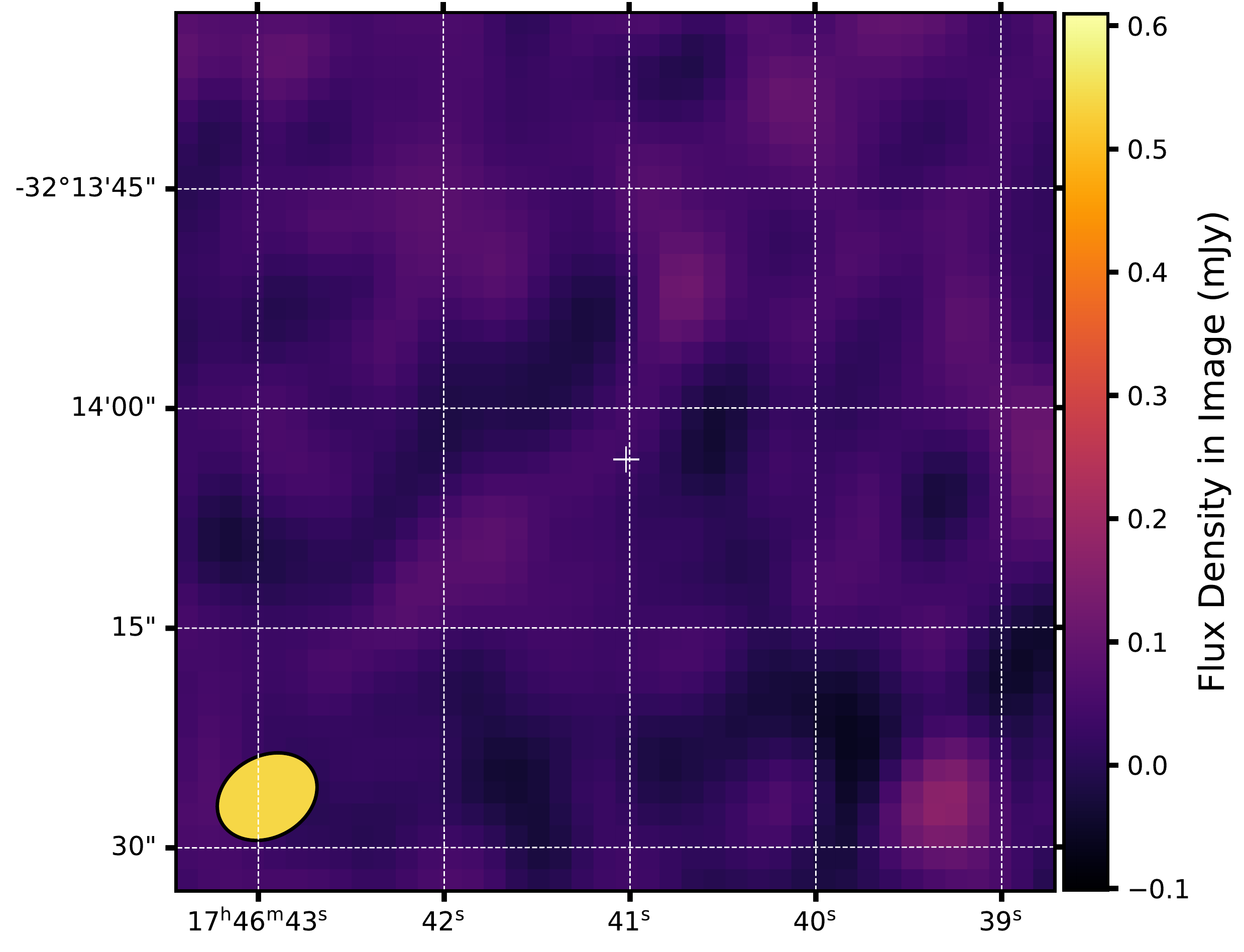}
	\includegraphics[width=.33\textwidth]{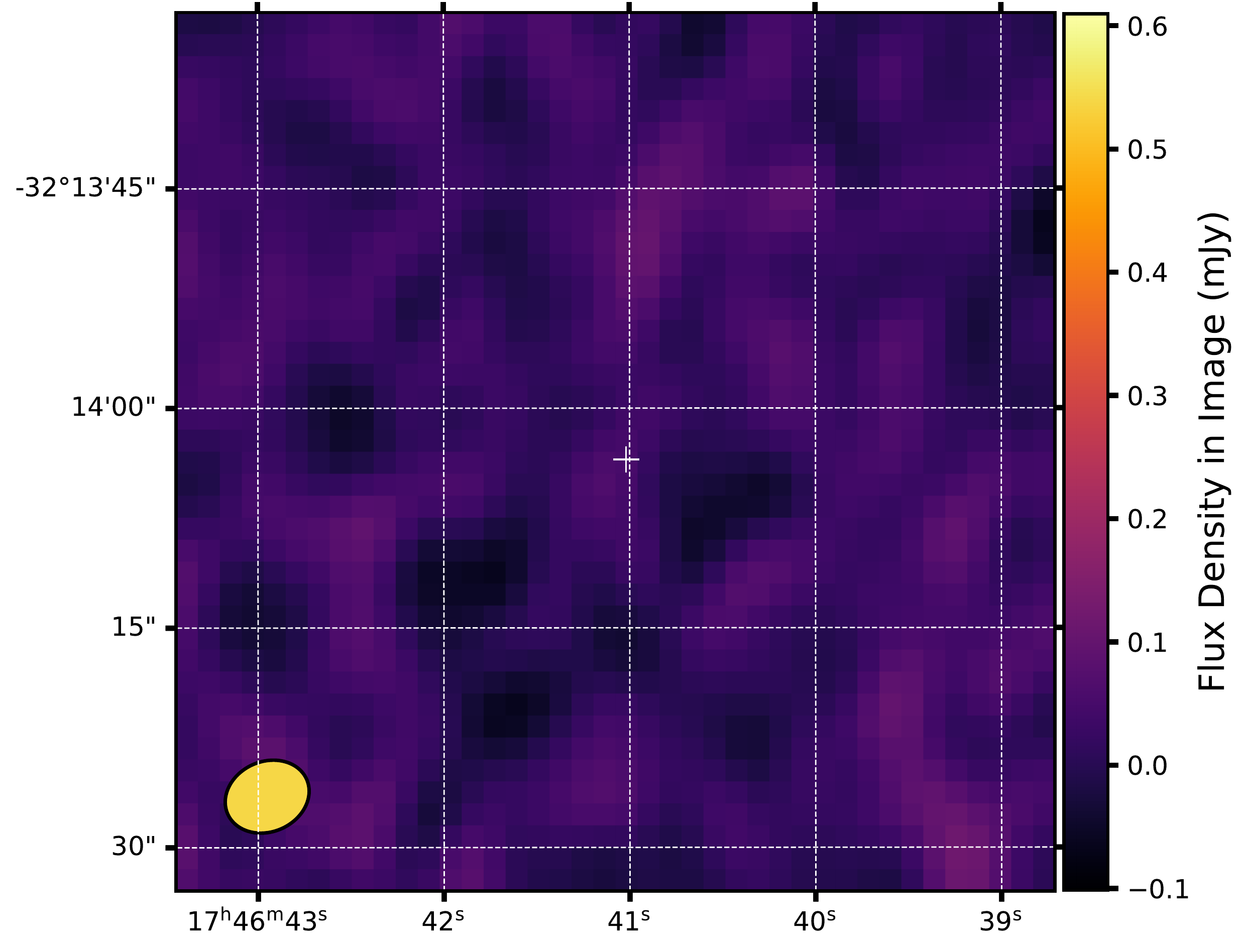}
	\includegraphics[width=.33\textwidth]{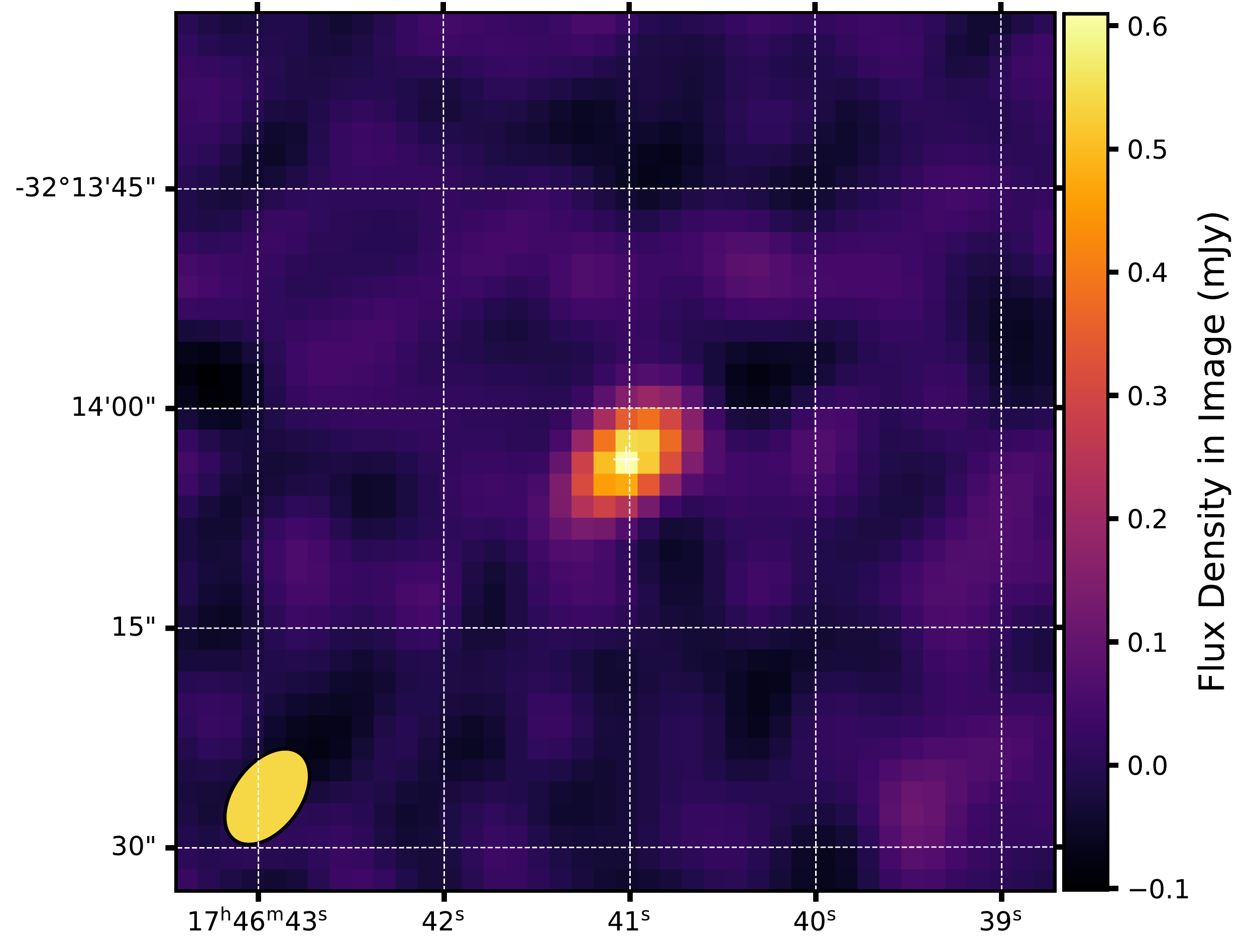}
	\includegraphics[width=.33\textwidth]{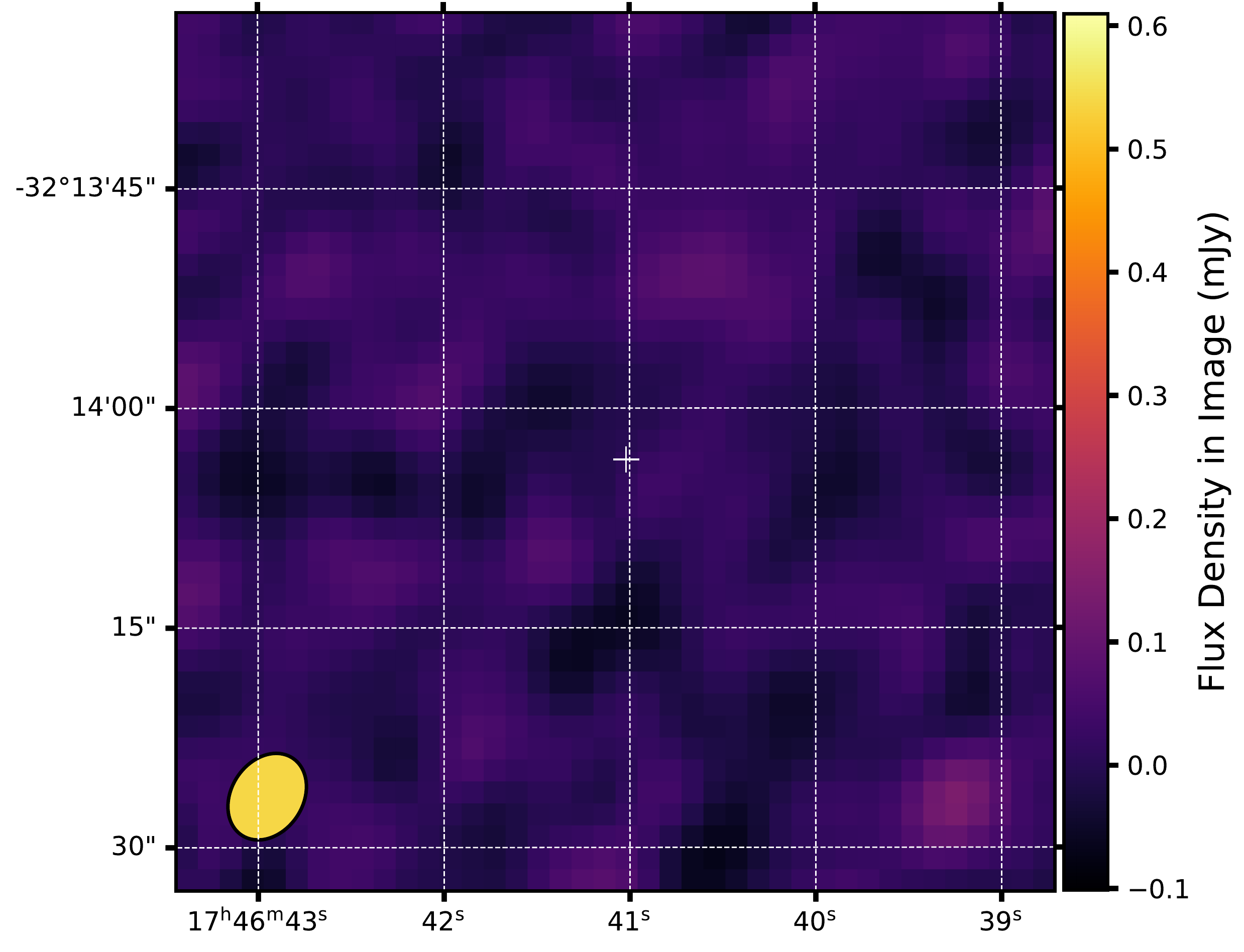}
	\includegraphics[width=.33\textwidth]{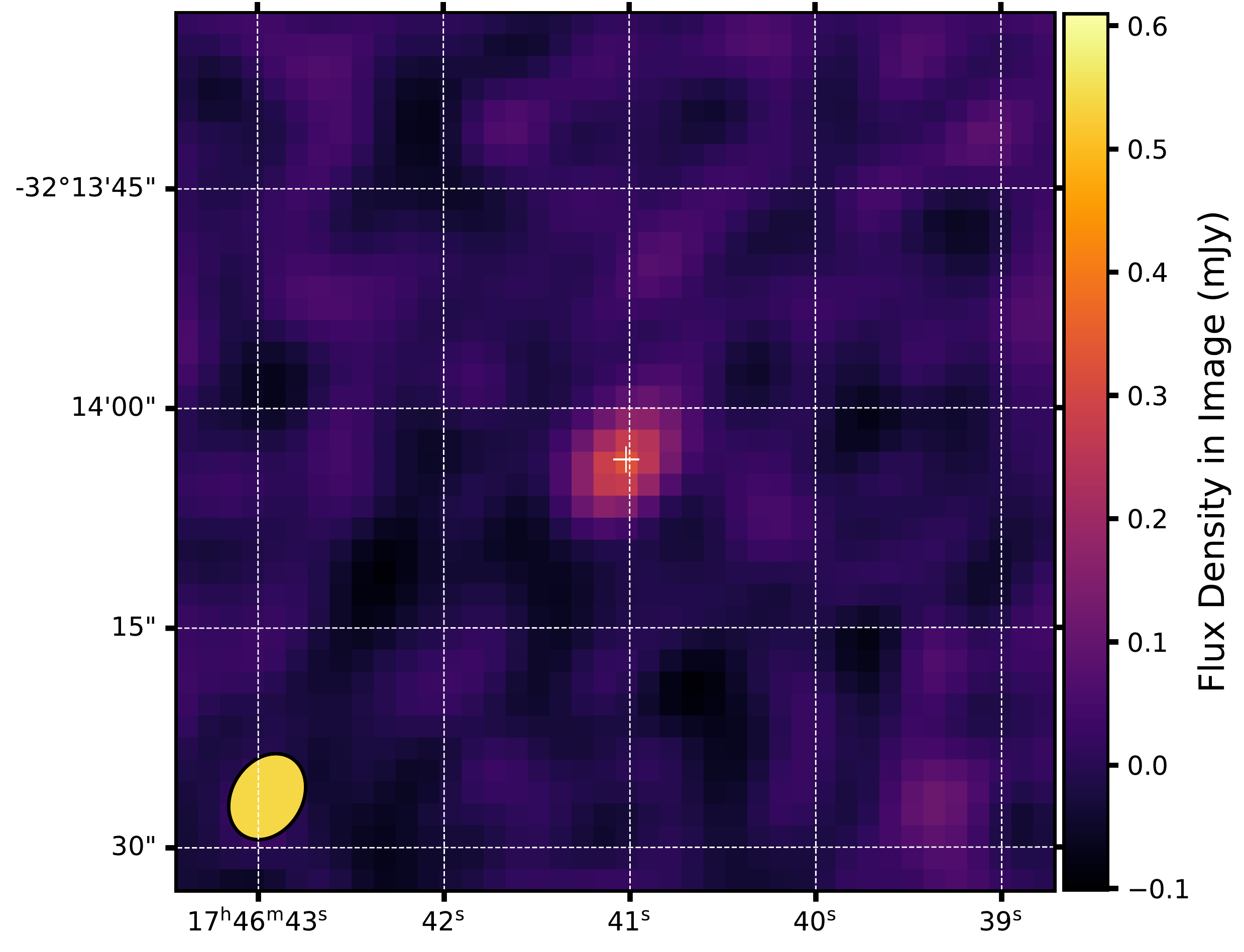}
	\includegraphics[width=.33\textwidth]{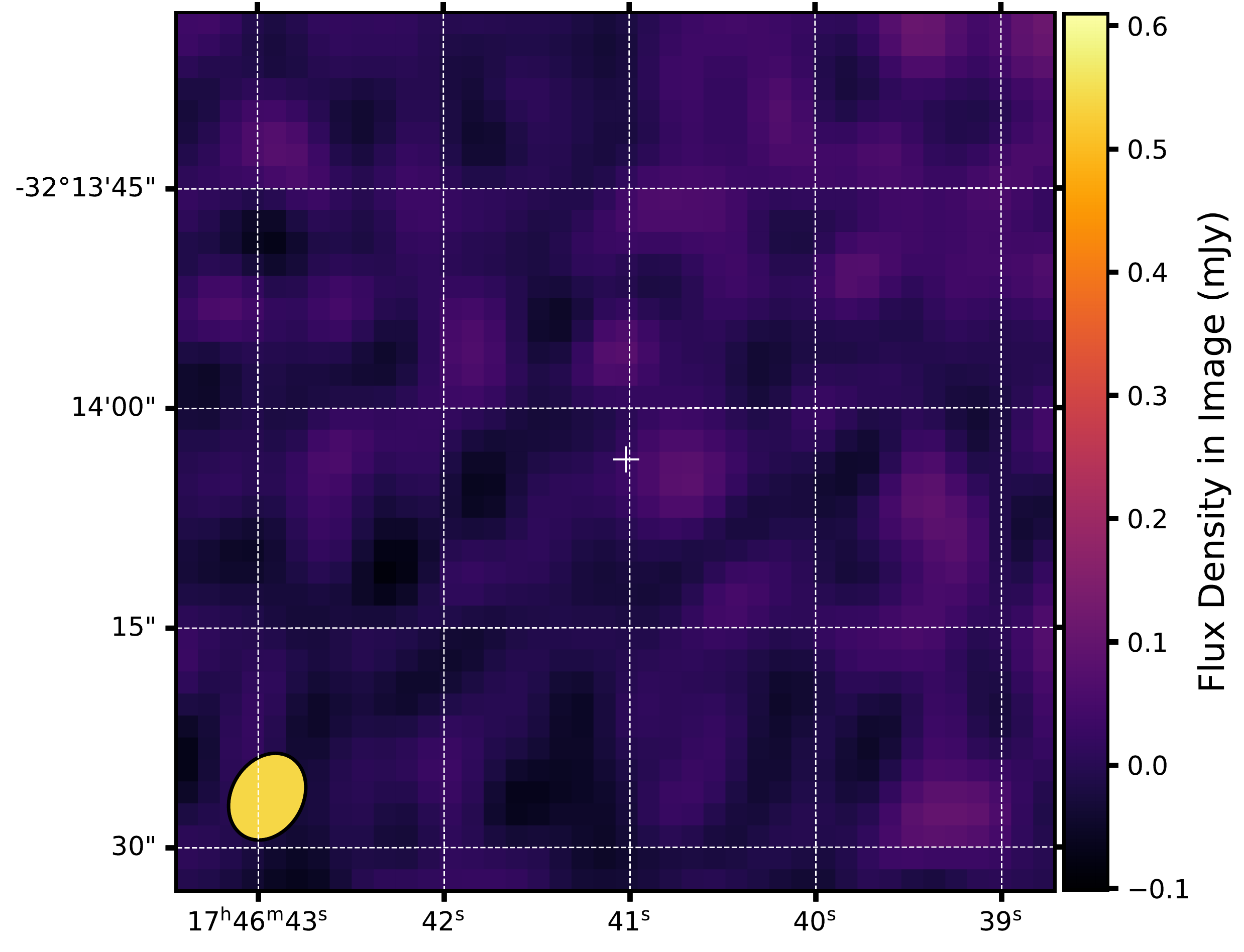}
    \caption{One arcminute square images of MKT J174641.0$-$321404 and surrounding noise during all ThunderKAT observations, running chronologically from top left, across then down. The source can be clearly seen to appear and disappear multiple times across all epochs. The synthesised beam of each image can be seen in the bottom left of each panel, whilst the white central cross-hairs show the source's mean MeerKAT position.}
    \label{fig:images}
\end{figure*}

\begin{figure}
	\includegraphics[width=\columnwidth]{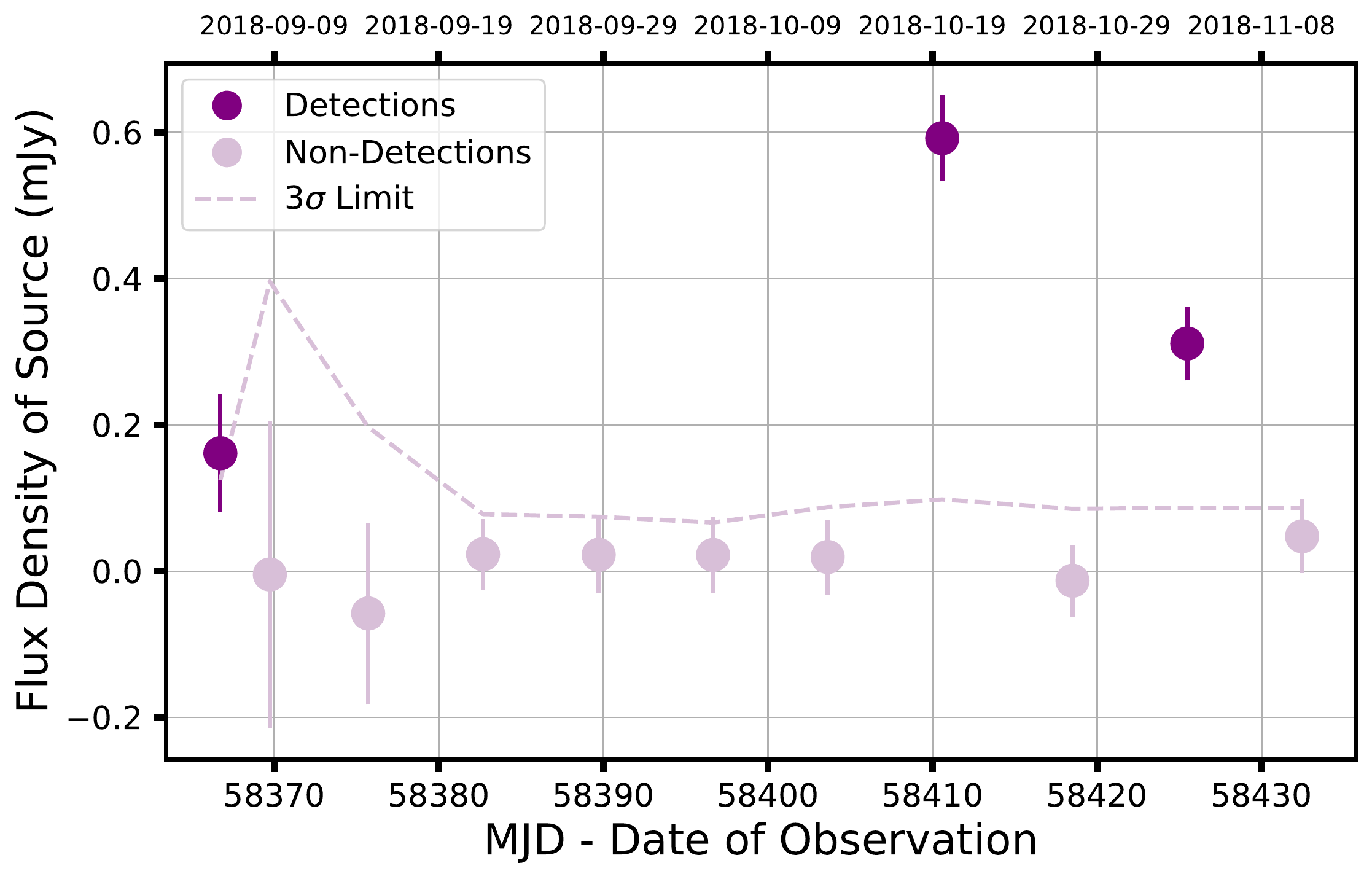}
    \includegraphics[width=\columnwidth]{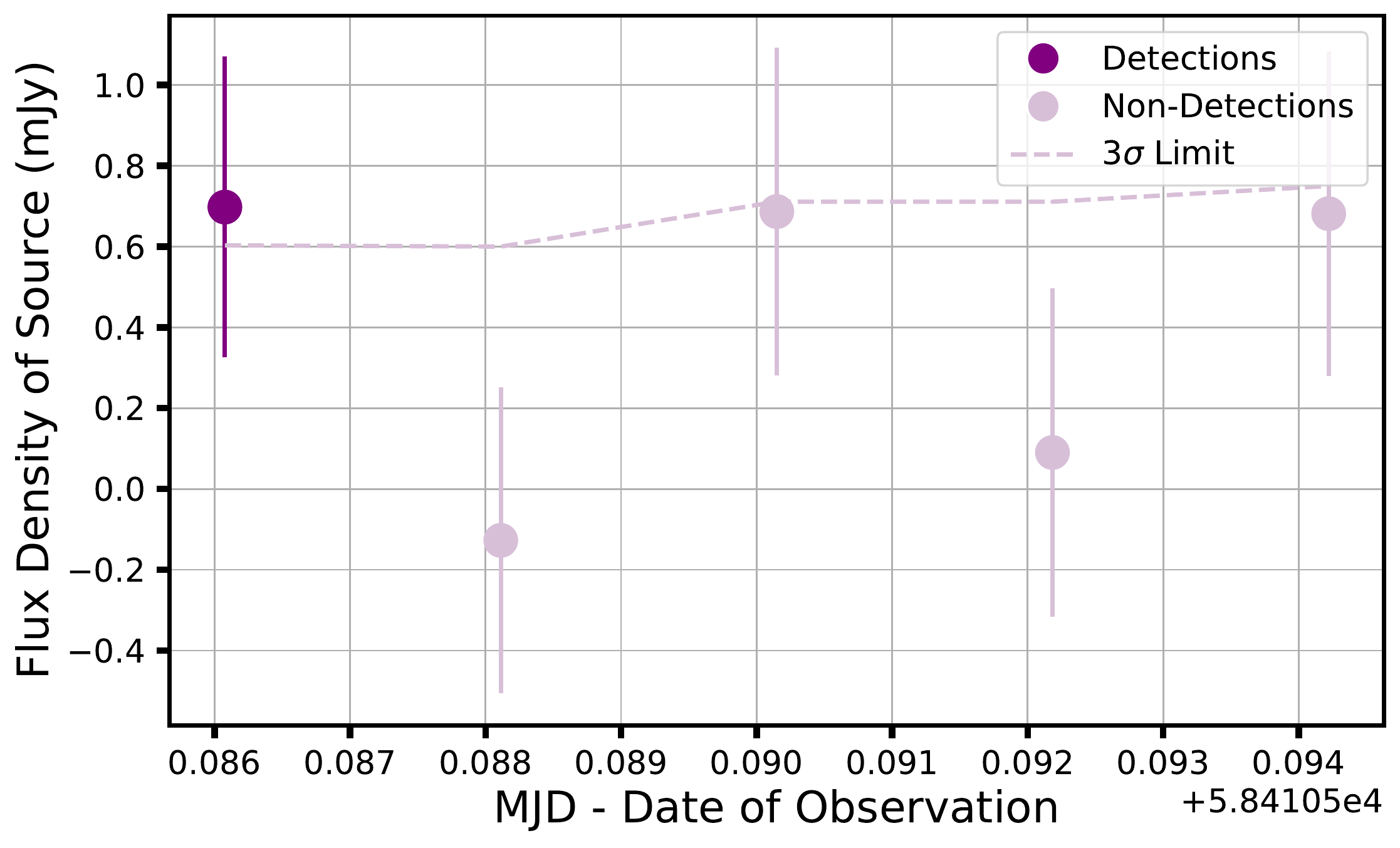}
    \caption{\textbf{Upper:} Radio light curve of MKT J174641.0$-$321404 across all 11 ThunderKAT observation epochs. Non-detections are defined as measurements below a 3$\sigma$ threshold (dashed line) where $\sigma$ is the local RMS noise in that epoch. The non-physical measurements in some epochs are caused by the forced photometry and epoch-specific noise structure, visible in Figure \ref{fig:images}. The higher detection threshold (due to more noise) in the first three epochs is likely due to the use of a different phase calibrator than the later observations.
    \textbf{Lower:} Radio light curve compiled from 5$\times$3 minute slices of the brightest 15 minute epoch. From this we can see evidence for marginal variability within the epoch.}
    \label{fig:LC}
\end{figure}

 This previously unknown radio source was detected serendipitously twice during the final four epochs of observation, at a maximum of 590 $\pm$ 60 $\mu$Jy  and can be seen to appear and disappear in the radio images of Figure~\ref{fig:images}. Running \texttt{TraP} whilst monitoring the source's location during every time step allows us to better understand the emission from MKT J174641.0$-$321404 and provide better constraints on its variability. The resulting light curve from such forced photometry of the source over all epochs can be seen in Figure~\ref{fig:LC}'s upper panel, where non-detections (defined as flux measurements detected at $<3\sigma$ above local noise) are indicated with faint points. The source can be seen to be detected during the first epoch of observation, at $\sim3.5\sigma$. Figure~\ref{fig:LC}'s lower panel shows marginal evidence for intra-epoch variability, wherein the brightest detection of MKT J174641.0$-$321404 has been split into 5 3-minute integrations. At $\sim 5'$ removed from the phase-centre of the radio images, the flux calibrations for this source are imperfect. However, the true values will not be so dissimilar as to invalidate our findings, especially as we are primarily interested in relative variations, and so no further observations have been made nor primary beam corrections applied to correct the light curve values. 

\section{Archival Multiwavelength Association}
\label{sec:Multi}

\begin{figure*}
\centering
	\includegraphics[width=0.8\textwidth]{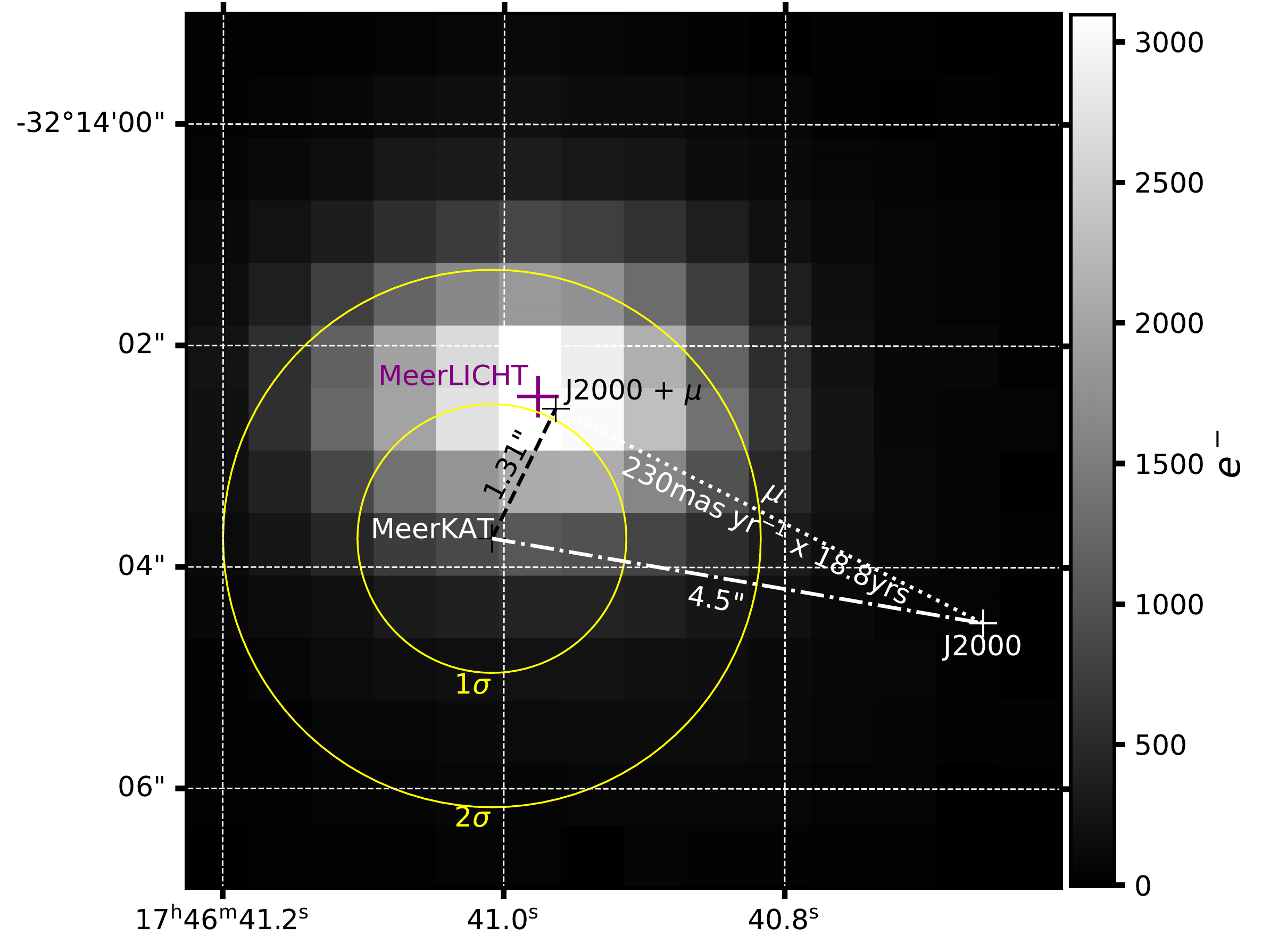}
    \caption{MeerLICHT q band (440-720 nm) reference image (pixels) of the region near radio source MKT J174641.0$-$321404. 1 and 2$\sigma$ positional uncertainties on the MeerKAT position (circles) show an overlap between the MeerLICHT object and the \textit{Gaia} proper motion attributed to SCR 1746.}
    \label{fig:optical}
\end{figure*}

The radio position of the source is taken to be the weighted mean of the three detection epochs using the Python Blob Detector and Source Finder \citep[PyBDSF;][]{Mohan2015}, differing slightly from \texttt{TraP} as the former includes the first epoch’s detection. At the time of research, cross-matching this mean position of $17^\mathrm{h}46^\mathrm{m}41\fs0, -32\degr14\arcmin04\arcsec  $(266$\fdg6709, -32\fdg2343 $) with the SIMBAD database\footnote{Set of Identifications, Measurements, and Bibliographies for Astronomical Data, accessed via \url{http://simbad.u-strasbg.fr/simbad/}. At time of first research in spring 2021, only SCR 1746-3214 was returned when cross-matching at this search radius.} returns two objects within $2\arcmin$, the optical sources SCR 1746$-$3214 \citep[henceforth SCR 1746, detection detailed in][]{Boyd2011} and \textit{Gaia} DR2 4055567846152508160. The latter is $\sim97\arcsec$ from the radio emission and so is likely too distant to be the origin thereof. SCR 1746 has an ICRS J2000 position of $17^\mathrm{h}46^\mathrm{m}40\fs659 -32\degr14\arcmin04\farcs50$ with uncertainties of $0.10$ and $0.8$ mas
respectively, calculated from \textit{Gaia} data release 2 (\citealp{Collaboration2016}; \citealp{Collaboration2018}), placing it at an on-sky distance from the radio coordinates of $4\farcs5$. SCR 1746 has a \textit{Gaia}-calculated distance from the Sun of 12.06$^{+0.1}_{-0.2}$ pc 
\citep{Bailer-Jones2018}.
Taking into account this star's high proper motion $\mu$ of (205.4 $\pm$ 0.2, 103.0 $\pm$ 0.1) 
mas yr$^{-1}$ in Right Ascension and Declination respectively, this star is $1 \farcs 3$ from the radio source. This is within the 2$ \sigma $ positional uncertainty of the radio position where $\sigma = $ 1$\farcs2$ and so we associate the radio transient events of MKT J174641.0$-$321404 as originating from SCR 1746. The next nearest optical source in the q-band MeerLICHT image (see below and Figure \ref{fig:optical}) is $ > $20 $ \arcsec  $ away.

Multiwavelength coverage from MeerLICHT \citep{Bloemen2016} confirms this association. 
A stack of q-band (440-720 nm) exposures taken on 7th June, 9th July and 16th July 2019 and 28th July 2020 form the MeerLICHT reference image for this field, where the small temporal gap between the radio and optical observations renders proper motion effects negligible.
The region surrounding MKT J174641.0$-$321404 can be seen in Figure~\ref{fig:optical}, showing the proper motion of the \textit{Gaia} source to within positional uncertainties of both the MeerKAT and MeerLICHT objects. This confirms that the radio transient coincides with SCR 1746.
We note that there are no observations simultaneous with the radio data as the strict MeerKAT-MeerLICHT coupling was not fully in-place during 2018, early in the life of the projects. In contemporary observations MeerLICHT follows MeerKAT pointings when possible, though the opposite is not true, hence there being no ThunderKAT data taken during MeerLICHT observations. 

\subsection{Optical and near-IR Photometry}

There is substantial photometric sky survey coverage of SCR 1746. Catalogued measurements of this source include the Two-Micron All-Sky Survey (2MASS; \citealp{Cutri2003}; \citealp{Skrutskie2006}) the AllWISE data release \citep{Cutri2013}, US Naval Observatory B catalog \citep[USNO-B;][]{Monet2003}, GLIMPSE Source Catalog \citep{SpitzerScienceCenter2009} and the optical SuperCOSMOS-RECONS (SCR) southern sky proper motion searches \citep{Boyd2011}. A non-exhaustive summary of some of the recorded magnitudes of this source can be found in Table~\ref{tab:fluxes} and is intended to illustrate the source's photometric properties of being relatively bright and red.

\begin{table}
	\centering
	\caption{Select magnitudes of red dwarf SCR 1746$-$3214 across the visible and near-IR spectrum. Bands are listed with their effective wavelength $\lambda_{\mathrm{eff}}$.}
	\label{tab:fluxes}
	\begin{tabular}{cccc}
		\hline
		 Band ($\lambda_{\mathrm{eff}}$) & Magnitude & Uncertainty & Reference\\
		  $\micron$                      & mag       & mag & \\
		\hline
		B (0.45)& 17.97 & 0.04 & \cite{Page2012} \\
		V (0.55)& 15.6 & 0.2 & \cite{Stassun2019} \\
		G (0.62)& 13.9734 & 0.0006 & \cite{Collaboration2018} \\
		I (0.88)& 12.98 & 0.30 & \cite{Sebastian2021} \\
		J (1.25)& 10.35 & 0.02 & \cite{Cutri2003} \\
		H (1.65)& 9.74 & 0.03 & " \\
		K (2.16)& 9.38 & 0.02 & " \\
		W1 (3.35)& 9.18 & 0.03 & \cite{Cutri2013}\\
		W2 (4.6) & 9.07 & 0.3 & "\\
		W3 (11.6)& 8.82 & 0.10 & " \\
		W4 (22.1)& 8.1 & 0.4 & " \\
		\hline
	\end{tabular}
\end{table}

\begin{figure}
\centering
    \includegraphics[width=\columnwidth]{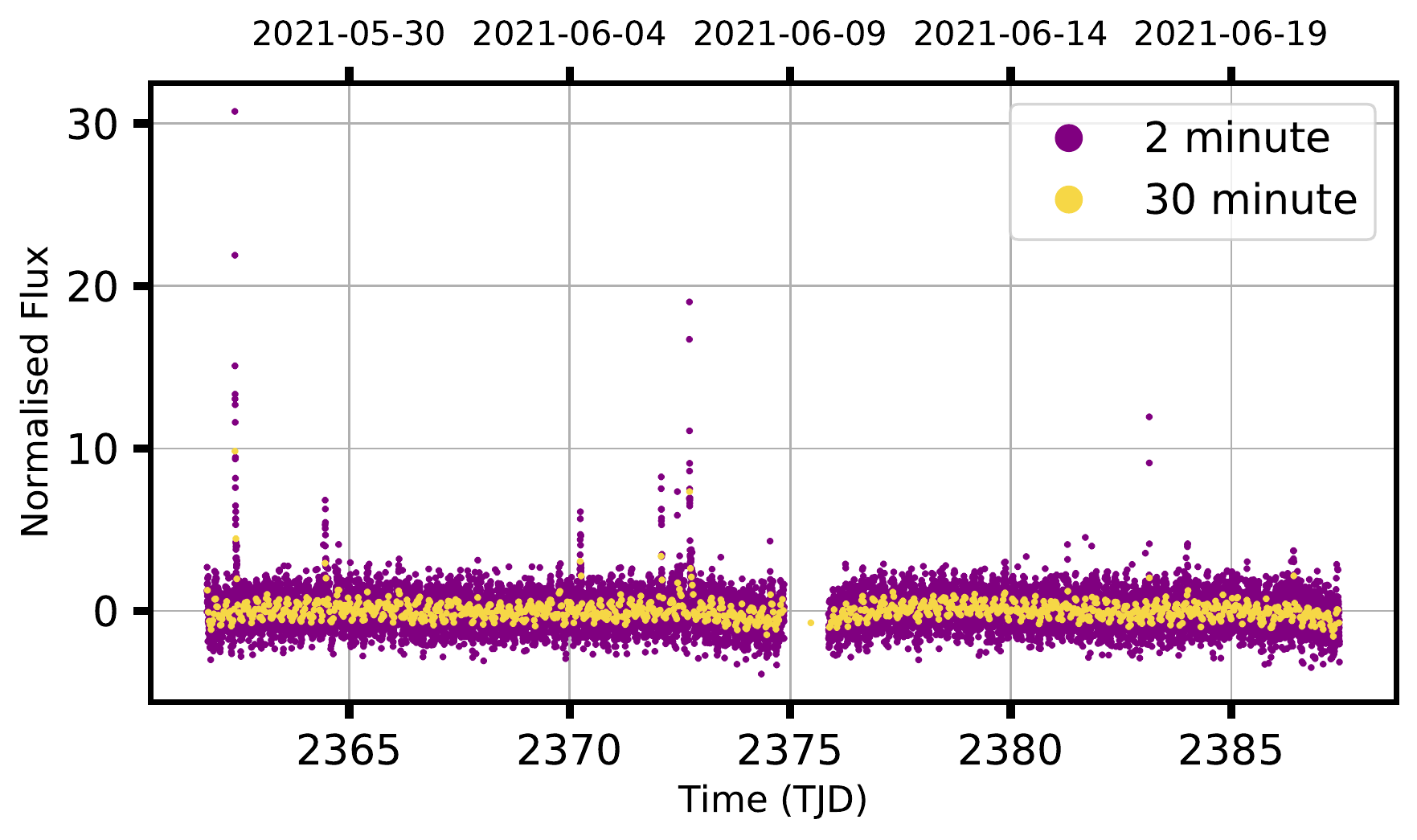}
    \caption{The \textit{TESS} light curve of SCR~1746 (Tess Input Candidate 111898820) spanning 27.9 days from May to June of 2021, including the 2-minute cadence data (purple) and the same data binned down to 30-minute cadence.$^{\ref{fnlabel}}$ Flare amplitudes can be seen up to $30\times$ above non-flaring times.}
    \label{fig:TESS}
\end{figure}

\begin{figure}
\centering
    \includegraphics[width=\columnwidth]{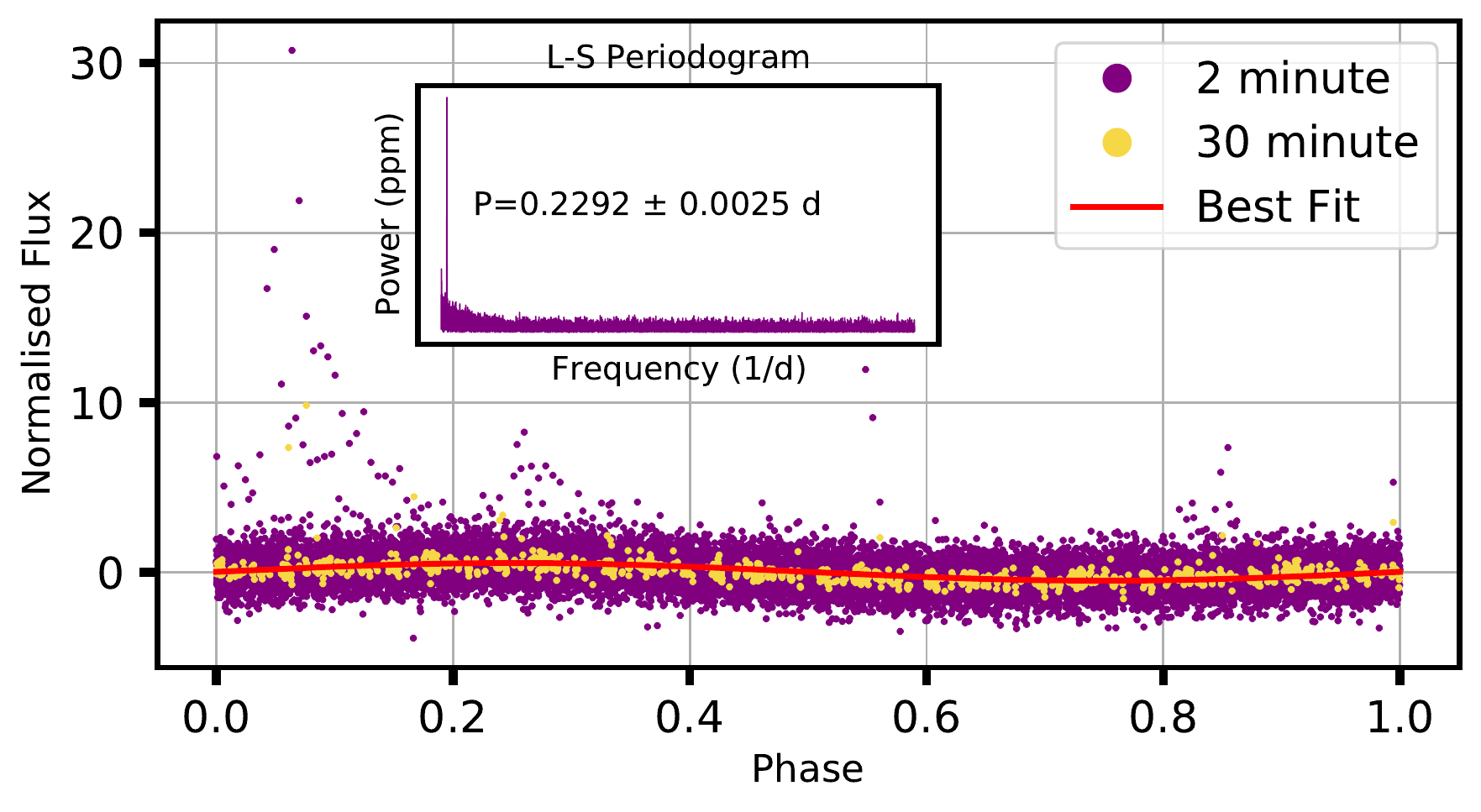}    \includegraphics[width=\columnwidth]{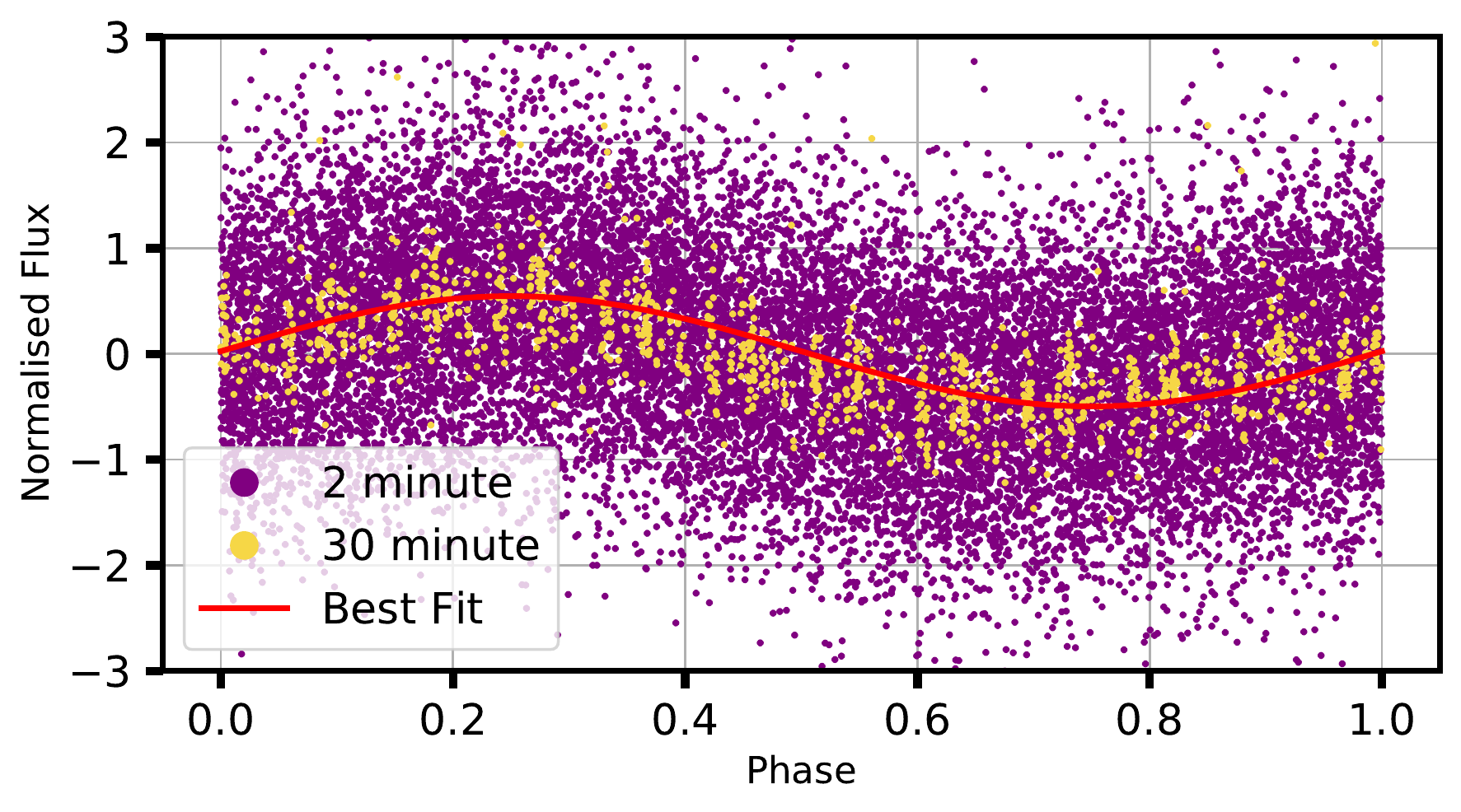}
    \caption{\textbf{Upper:} The phase folded \textit{TESS} light curve of SCR 1746, with the corresponding Lomb-Scargle periodogram inset. Flares tend to occur between 0.0 and 0.4 in phase, corresponding to the brighter parts of the quiescent light curve. A sinusoidal best fit to the unbinned data is plotted, showing a relative amplitude of 0.52~$\pm$~0.01.
    \textbf{Lower:} A closer view at the non-flare behaviour of the phase-folded light curve, in which the amplitude of the best fit sine curve is clearly visible. The 30-minute cadence data reduces the scatter and falls in good agreement with the sinusoid.}
    \label{fig:PhaseFold}
\end{figure}

SCR 1746 \citep[Tess Input Candidate 111898820; ][]{Stassun2019} was also observed by NASA's Transiting Exoplanet Survey Satellite \citep[\textit{TESS}; ][]{ricker15} during Sector 39 of the extended mission. The data, which span 27.9 days between 27 May and 24 June 2021, consist of observations obtained every 2-seconds. These images were combined into 2-minute cadence data products on board of the spacecraft prior to being processed and reduced by the Science Processing Operations Center \citep[SPOC; ][]{jenkinsSPOC2016}. The full \textit{TESS} data set, shown in Figure \ref{fig:TESS} binned to 2-minute and 30-minute cadence\footnotemark\footnotetext{We note that the chosen binning matches the \textit{TESS} Full Frame Image time-scale by coincidence.\label{fnlabel}}, shows clear flares that reach up to 30 times quiescent brightness levels. The data were searched using a Lomb-Scargle periodogram \citep{Lomb76, Scargle82} to identify periodic signals in the light curve, revealing a significant periodicity with a period of 0.2292~$\pm$~0.0025~days. When phase folding the \textit{TESS} light curve at this period, a sinusoidal fit to the data produces a best fit relative amplitude of 0.52 $\pm$ 0.01 (bottom panel of Figure \ref{fig:PhaseFold}). There are some background sources on the same \textit{TESS} pixel, but at G magnitudes of 18.9 or fainter (c.f. 13.95 for SCR~1746) we conclude that these are all too faint to be causing the observed modulation. As can be seen in the top panel of Figure \ref{fig:PhaseFold}, more flares tend to occur between phase 0 and 0.5, corresponding to the brightest parts of the phase curve. This is in keeping with the photometric modulation being caused by magnetically active spots and associated bright plages rotating into or out of view. If we assume that the star rotates as a rigid body with radius $R$ (see section \ref{Sect:PhotSpec}) we can express the photometric period as a tangential velocity $v_t = 2\pi R/P = 32.2\pm0.9~ $km~s$^{-1}$, demonstrating that SCR 1746 is a fast rotator, comparable to other low mass systems (c.f. \citealp{Gizis2017} for other dwarfs, or the solar value of $\sim$ 2 km s$^{-1}$).

\begin{figure}
    \centering
    \includegraphics[width=\columnwidth]{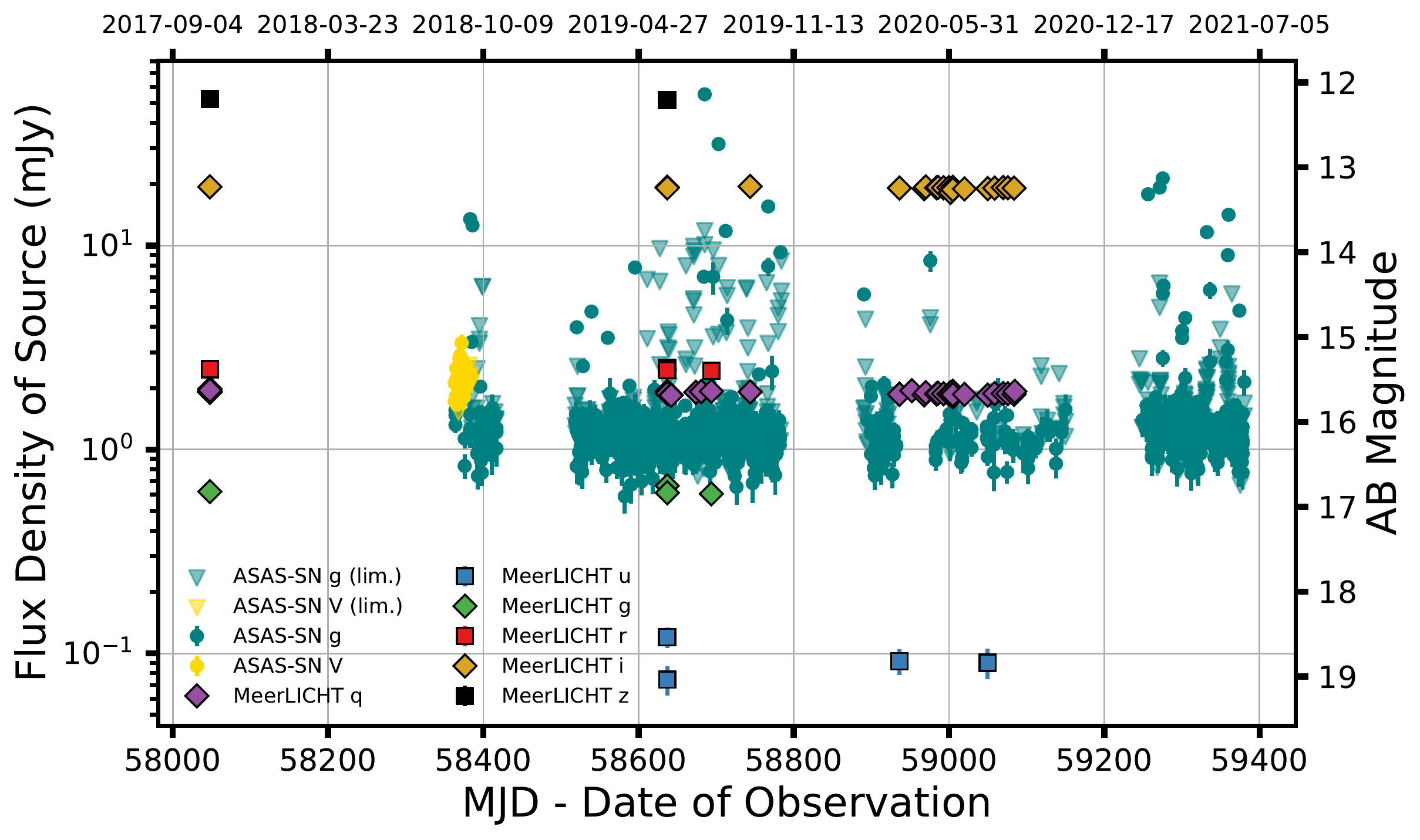}
    \includegraphics[width=\columnwidth]{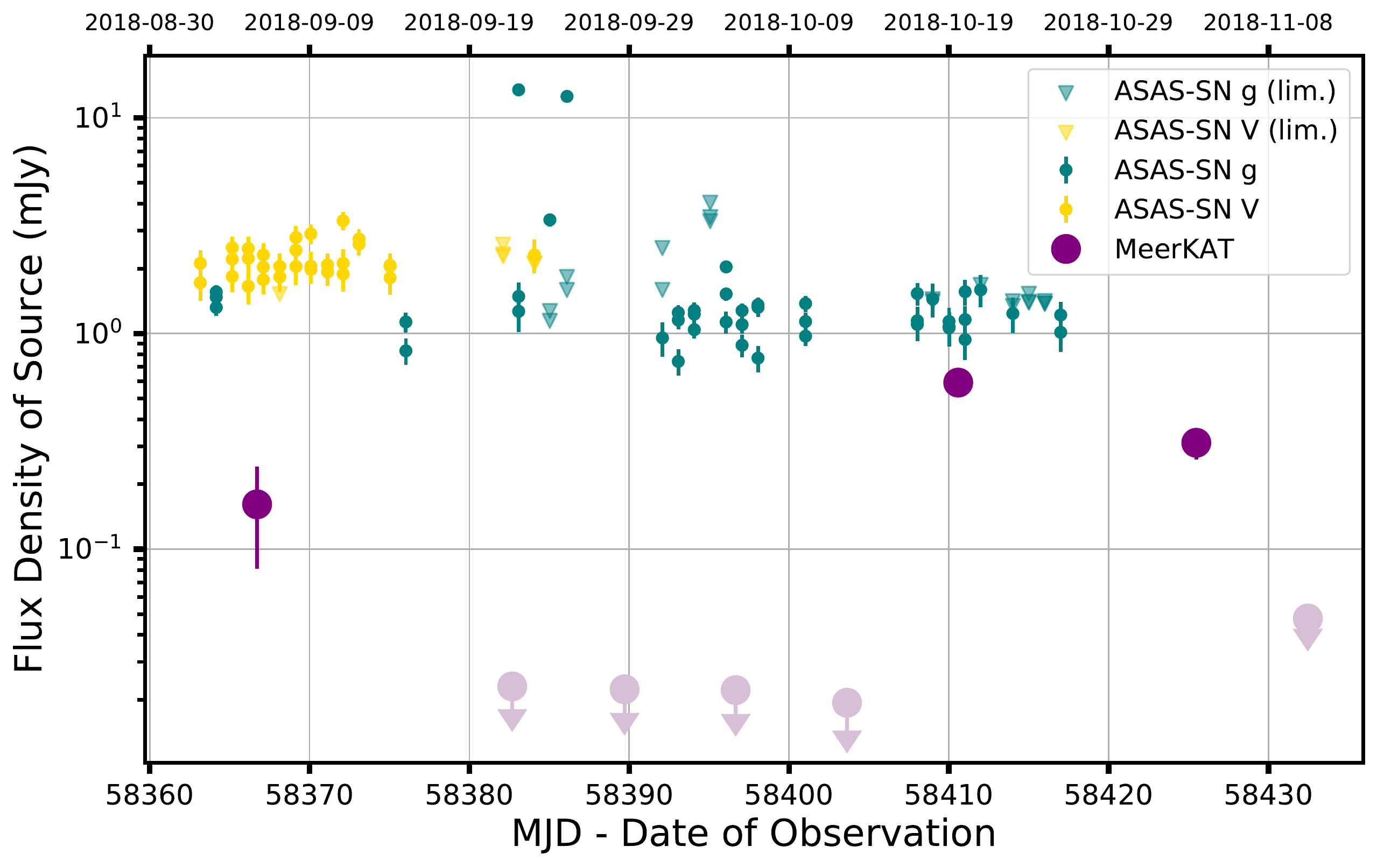}
    \caption{\textbf{Upper:} Optical data from MeerLICHT and ASAS-SN.  Flaring behaviour can be seen in data from the ASAS-SN Sky Patrol portal between radio observations and mid-2021. Downwards facing triangles indicate upper limits in their respective bands.
    \textbf{Lower:} The radio light curve of SCR 1746 overlaid with the simultaneous optical data. No flares are simultaneous to within $<0.4$ days with those in the other bands. }
    \label{fig:optLCs}
\end{figure}

Additional optical monitoring of the source is provided by both MeerLICHT (see Section \ref{sec:intro}) and the All-Sky Automated Survey for Supernovae \citep[ASAS-SN, a global network of 24 40-cm telescopes;][]{Shappee2014,Kochanek2017}.
The optical data in Figure~\ref{fig:optLCs} spans almost 4 years of observations in Johnson's V, Sloan ugriz and q-band filters. MeerLICHT rotates between filters every 2 minutes, allowing for quasi-simultaneous colour determination (see Section \ref{Sect:PhotSpec}). From the MeerLICHT data we can clearly see the star is much brighter at redder wavelengths.
Aperture photometry at a 5$\sigma$ detection threshold from the ASAS-SN Sky Patrol portal shows that optical flares are likely to have been observed from this source in the g-band. Despite this, none of the optical or radio flares are simultaneous to within $<0.4$ days. Given that flares can be as short as minutes, this is not unexpected behaviour. Similarly, the photometric period observed by \textit{TESS} is not observed in these data due to infrequent sampling with diurnal and seasonal gaps.

\subsubsection{Photometric spectral typing}
\label{Sect:PhotSpec}
The latest version of the \textit{TESS} Input Catalog \citep[TIC v8;][]{Stassun2019} estimates the target to have an effective temperature of 2870~$\pm$~160~K, a mass of 0.12 $\pm$ 0.02 $M_{\sun}$ and radius of 0.146 $\pm$ 0.004 $R_{\sun}$. \cite{Sebastian2021} compile a target list of ultracool dwarf systems in preparation for \textit{TESS}, wherein an empirical relationship for $T_{\mathrm{eff}}(M_H)$ from \cite{Filippazzo2015} is inverted to show that the target has a spectral type of M5.1. Their calculated temperature, mass and radius are all slightly different than, but consistent with, those from \cite{Stassun2019}. The effective temperature scale of M dwarfs given by \cite{Rajpurohit2013} also indicates that a temperature of $\sim2900$ K is in agreement with a mid-M spectral typing. Finally, MeerLICHT observations provide quasi-simultaneous colours $(r-i) = 2.22 \pm 0.02 $ and $(i-z) = 1.09 \pm 0.02$, in agreement with typical values for M6 stars \citep{2005PASP..117..706W, Douglas2014}.


\subsection{X-ray Observation and Correlation}

X-ray counts from the 7 year Swift-XRT point source catalog \citep[1SWXRT;][]{DElia2013}, taken over a total exposure time of 4631.1s  on the H1743$-$322 field are found to be coincident within positional uncertainties of SCR 1746. Assuming an absorbed power-law spectrum, the associated X-ray source has a 0.3-10keV flux upper limit of  $1.68 \times 10^{-13}$ mW m$^{-2}$, measured at $S/N = 2.8$. 1SWXRT has been superseded by the Swift-XRT Point Source catalogs 1 and 2 (1SXPS; \citealp{Evans2014}, 2SXPS; \citealp{Evans2020}), both of which contain nearby sources (separation in 1SXPS = 7$\farcs$5, 2SXPS = 2$\farcs$31), but that do not formally lie close enough to the GAIA position for association within a 90 per cent confidence interval region. As such, herein we use the analysis from 1SWXRT, but it is worth noting that the equivalent fluxes for the nearest sources in 1SXPS and 2SXPS respectively are $\sim1.16\times$ and $\sim0.43\times$ the 1SWXRT flux, so the overall conclusions made here would still broadly apply.

There is a known relation between the quiescent radio ($L_R$) and X-ray ($L_X$) luminosities of several types of active star, indicating  a connection between the nonthermal, energetic electrons causing the radio emission and the bulk coronal plasma responsible for thermal X-rays: 
\begin{equation}
    \mathrm{log}(L_X) \lesssim \mathrm{log}(L_R) + 15.5
    \label{eqn:GB}
\end{equation}
\citep{Guedel1993, Benz1994}. The standard interpretation of this Güdel and Benz relation is that magnetic reconnection in the corona accelerates a population of non-thermal electrons that emit at radio frequencies via gyrosynchrotron emission. These electrons also heat the chromosphere of the star, producing thermal X-ray emission. Note that coherent emission processes are known to violate this relationship, as shown by \cite{2021NatAs.tmp..196C}, making agreement with the Güdel and Benz relation an effective diagnostic of emission mechanism (see Section \ref{sec:Disc}).

\begin{figure}
\centering
	\includegraphics[width=\columnwidth]{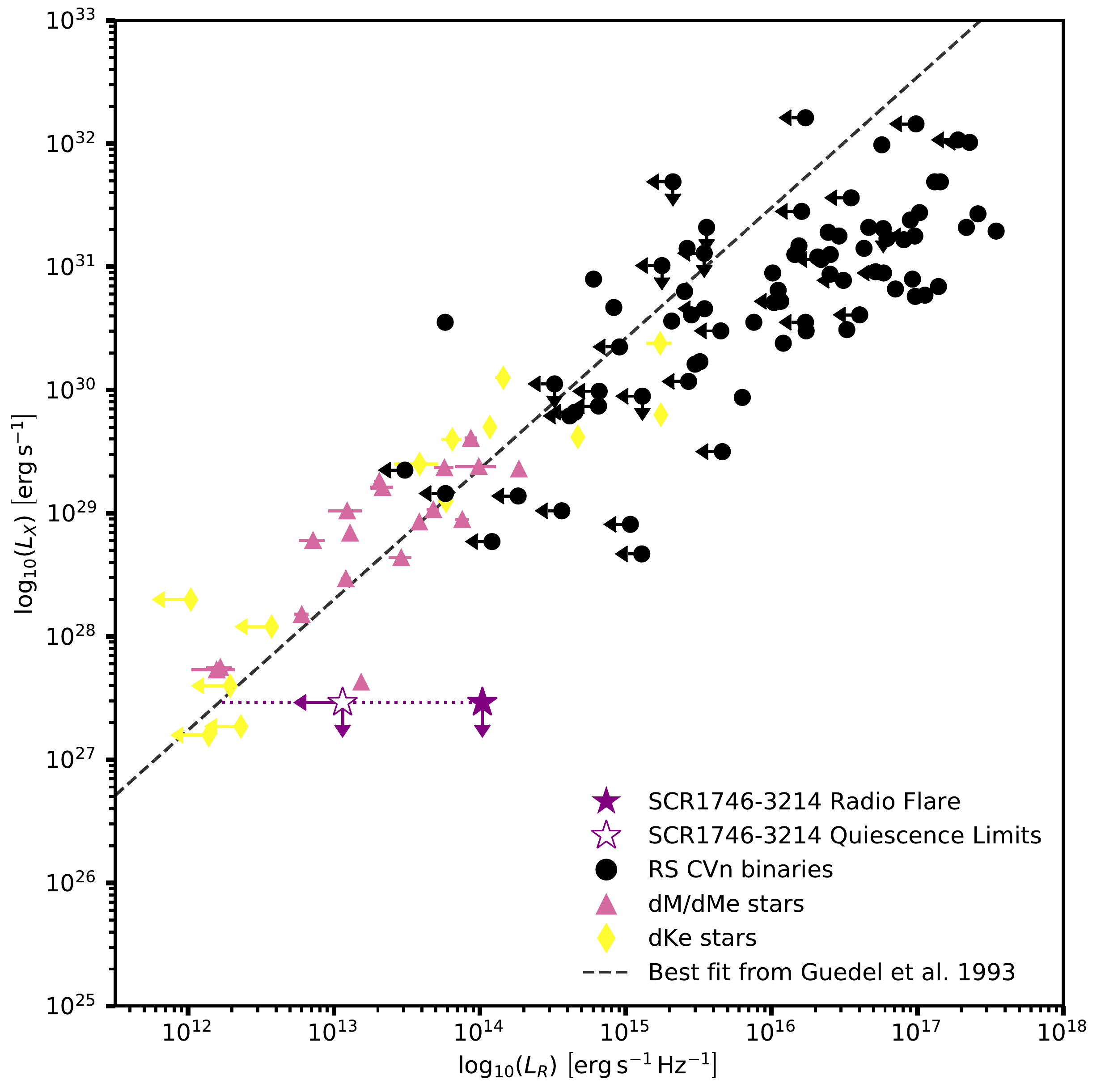}
    \caption{X-ray and radio luminosities of several types of active star \citep{Guedel1993} - reproduced from \url{https://github.com/AstroLaura/GuedelPlot} . Limits on SCR 1746's quiescent emission show this is an intrinsically faint source and could be approximately consistent with the Güdel and Benz relation for M dwarf (dM/dMe) stars. The radio flare is also plotted, showing a brightness increase of over an order of magnitude above quiescent limits.}
    \label{fig:GB}
\end{figure}

Figure~\ref{fig:GB} shows the quiescent limits and flare magnitude of SCR~1746 in comparison to other M dwarfs and other types of active star. This figure shows that M- and K-type dwarfs, some suffixed `e' to denote emission lines (dM/dMe and dKe, pink triangles and yellow diamonds respectively) tend to have lower radio and X-ray luminosities than the RS CVn binaries denoted by black circles. The quiescent radio limit for SCR 1746 is 3$\times$ the lowest RMS noise floor across the individual MeerKAT epochs, measured locally using PyBDSF to be $\sim 22 \mu$Jy. This corresponds to a specific luminosity upper limit of 1$\times10^{13}$erg s$^{-1} $Hz$^{-1}$, an order of magnitude below the brightest observed flare at $(1.0 \pm 0.1) \times10^{14}$ erg s$^{-1}$ Hz$^{-1}$. We note that these are only approximate positions of the SCR 1746 in the figure, as the quiescent radio and X-ray emission measurements should be taken simultaneously, which has not been the case here.

\section{Optical Spectroscopy}
\label{sec:Spect}
SCR 1746 was observed with the 11-m Southern African Large Telescope \citep[SALT;][]{Buckley2006} to provide confirmation of spectral type and investigate the magnetic activity of the star.
Two consecutive 1000s exposures were taken using SALT's High Resolution Spectrograph \citep[HRS;][]{Bramall2012, Crause2014} starting at UT 23:49:05 on the 11th of July 2021 at a seeing of $\sim 2\farcs0$. The HRS is a high dispersion échelle spectrograph and was operated in low-resolution mode at a spectral resolving power of $\sim$16500. Wavelength calibration is performed relative to ThAr arc spectra on a weekly basis and the data were reduced accounting for trimming, bias subtraction, gain correction, cosmic ray cleaning and flat fielding, done via PySALT \footnote{\url{http://pysalt.salt.ac.za/}} \citep{Crawford2016}. Whilst HRS is dual beam, providing both blue (370-555nm) and red (555-890nm) spectra, as archival photometry indicates SCR 1746 is highly red, we opt to only analyse the latter.

\protect{
\begin{figure}
\centering
	\includegraphics[width=\columnwidth]{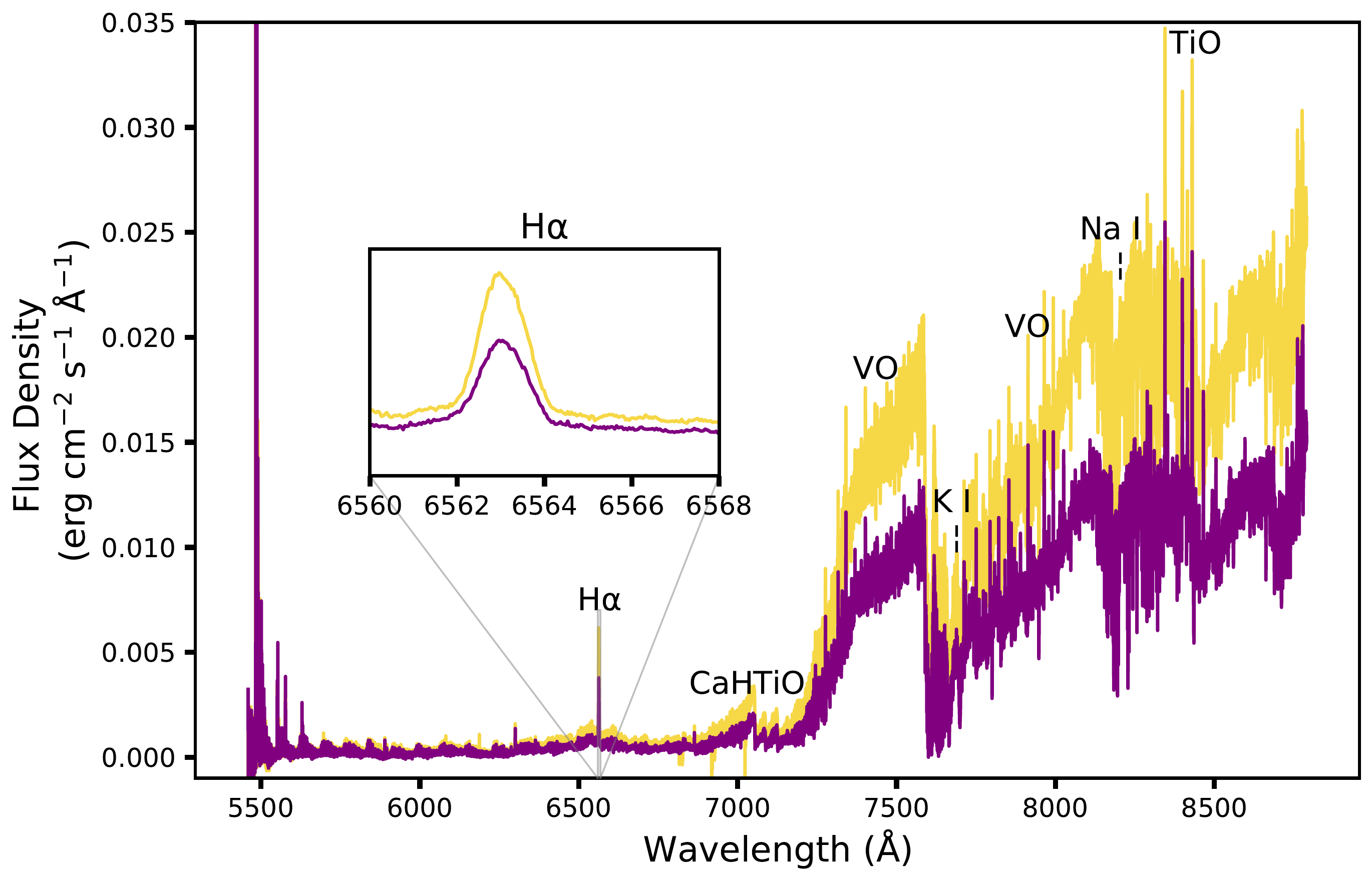}
	
\caption[Caption with citation]{ Two reduced spectra of SCR 1746$-$3214, taken by SALT. Some notable spectral features have been indicated, including clear H$\alpha$ emission and molecular lines, estimated from ~\cite{Kirkpatrick1991}. Inset is a view of the H$\alpha$ emission line, which shows rotational broadening (see Section \ref{sec:Disc}).}
	
    \label{fig:spectra}
    
\end{figure}
}

The two red spectra taken of SCR 1746 can be seen in Figure~\ref{fig:spectra}. The extensive structure at wavelengths >7000 Å is typical for late M dwarfs and is caused by absorption due to molecules such as TiO, whilst H$\alpha$ can also be seen in emission at $\sim$ 6560 Å. Following the prescription of \cite{Newton2017}, the equivalent widths (EW) of the H$\alpha$ emission line integrated between 6558.8 Å and 6566.8 Å were calculated relative to a continuum level either side of the feature, between 6500–6550 Å and 6575–6625 Å. The mean EW and its standard deviation of the two spectra are
$-5.5 \pm 0.2$ Å placing SCR 1746 clearly in the population of magnetically active stars (EW $\leq -1$ Å; see \citealp{Newton2017}).

In order to estimate a spectroscopic classification and metallicity for SCR 1746, data were compared to co-added templates of individual stellar spectra from the Sloan Digital Sky Survey’s Baryon Oscillation Spectroscopic Survey \citep[SDSS BOSS;][]{Dawson2013} using the spectral typing code \texttt{PyHammer} \citep{Kesseli2017, Roulston2020}\footnote{\url{https://github.com/BU-hammerTeam/PyHammer}}. The weighted mean and variance of 34 spectral indices from  all co-added spectra across spectral types O5 through L3 and metallicities $-$2.0 $<$[Fe/H] $<$ +1.0 dex provide the empirical models against which chi-squared minimisation is performed relative to the same features in the input spectra. The model of smallest chi-squared is therefore the closest to the input spectra and visual inspection allows for confirmation of this. Using this analysis, the best fitting empirical template for both SALT spectra is of an M8 star with metallicty $Z = -0.5$. We can take the standard deviations of the \texttt{PyHammer} results when tested on their own SDSS BOSS spectra i.e. those used to build the empirical templates, as pessimistic estimates for uncertainties in classification and metallicity $Z$, which come to $\pm$1.5 spectral subtypes and $\pm$0.4 dex  respectively (see the Appendix of \citealp{Kesseli2017} for more on this.).

\section{Discussion}
\label{sec:Disc}


Commensal analysis of MeerKAT images of the sky surrounding H1743$-$322 identified a new radio transient, MKT J174641.0$-$321404, coincident with the high proper motion star SCR 1746$-$3214. The radio flaring seen in late 2018 was detected three times over 11 epochs of data from MeerKAT. This marks the second serendipitous, galactic transient discovered by the ThunderKAT team, both associated with kinds of active star. SCR 1746 is also the second M dwarf detected by MeerKAT \citep[see likely quiescent emission from targeted searches of ThunderKAT commensal data in][]{Driessen2021}. Being only a few arcminutes from the black hole XRB H1743$-$322, more commensal observations of the flare star are likely if the XRB enters an outburst phase over the coming few years.

The brightness temperature $T_b$ of a source is a useful diagnostic of emission mechanism. For radio emission of flux density $f_{\nu}$ observed at a distance $d$ and frequency $\nu$ 

\begin{equation}
    T_b = 2\times10^9 \left(\frac{ f_{\nu}}{\mathrm{mJy}}\right) \left(\frac{\nu}{\mathrm{GHz}}\right)^{-2} \left(\frac{d}{\mathrm{pc}}\right)^2 \left(\frac{ L}{R_{\mathrm{Jup}}}\right)^{-2}  \mathrm{K}
    \label{eqn:Tb}
\end{equation}
\citep{Dulk1985,Burgasser2005}, where $R_{\mathrm{Jup}}$ is the radius of Jupiter. We take length scale $L$ to be 2 stellar radii (radius $\sim 0.146 R_{\sun} $ from the TIC; \citealp{Stassun2019}), the approximate size of M dwarf corona \citep{Benz1995}. For the brightest radio detection this produces a brightness temperature of $\sim 10^{10}$ K. We can relax the assumption of a length scale by taking the variation on time-scales $\sim$3 minutes (see the lower panel of Figure \ref{fig:LC}) to correspond to a maximum emitting region. Doing so produces a minimum brightness temperature of $9\times10^{7}$ K. This is below the expected $10^{12}$ K maximum for incoherent emission and so the mechanism could be either incoherent or coherent in nature. The clearest evidence for a coherent mechanism would be the presence of high degrees of circular polarisation, a measurement we do not currently have as the appropriate calibration for short-integration MeerKAT data not at phase-centre are still being developed. A factor of $\gtrsim$10 brightness increase on quiescent emission lasting for time-scales of at least minutes is also in keeping with incoherent radio bursts \citep{Osten2007}. Futhermore, the magnetic activity seen by the measured H$\alpha$ EW and the approximate position of SCR 1746 on the Güdel and Benz relation (Figure \ref{fig:GB}) suggest that gyrosynchrotron emission is responsible for the observed emission. Further radio observations such as polarimetry, broader frequency coverage to constrain the spectral index, or simply a longer observation campaign will help determine the exact nature and frequency of the radio flares.

Optical photometry clearly indicates SCR 1746 is bright and red, with photometric relations suggesting a mid-M spectral type. This is in mild disagreement with the findings from the SALT spectrograph when run through \texttt{PyHammer}, which indicates a later spectral type of M8 (see Section \ref{sec:Spect} and Figure \ref{fig:spectra} therein), thus suggesting it falls under the ultracool dwarf designation. The properties of ultracool dwarfs beyond spectral type M7 depart from expectations set by earlier types \citep{Berger2002,Burgasser2005,Berger2010}, in the most extreme cases deviating from the Güdel and Benz relation (plotted in Figure \ref{fig:GB}) by four orders of magnitude (e.g. \citealp{Berger2001}). These ultracool dwarfs show a marked decrease in magnetic activity (seen in reduced relative H$\alpha$ flux) and relative X-ray flux, but an almost constant level of radio emission, i.e. the relevant coupling between wavebands appears to no longer hold in ultracool dwarfs. For example, \cite{2021NatAs.tmp..196C} demonstrate that radio emission of 19 M dwarfs detected at 144 MHz deviate from the Güdel and Benz relation and are caused by coherent processes i.e. different to that detected for late-type stars at gigahertz frequencies and not related to their chromospheric activity. To provide a relative H$\alpha$ to bolometric flux density we employ the $\chi$ method of \cite{Walkowicz2004}. Calculating $\chi$ from SCR 1746's $(i-J)$ colour \citep{Douglas2014}, then $L_{\mathrm{H\alpha}}/L_{\mathrm{bol}} = \mathrm{EW_{H\alpha} \times \chi} = (8.1 \pm 2.5)\times 10^{-5} $. This relatively strong H$\alpha$ flux is again in agreement with a spectral type M5-M6, before the breakdown of the radio--X-ray--H$\alpha$ couplings (see Figure 5 of \citealp{Berger2010}). Furthermore, the spectral indices of late M dwarfs are not well characterised - \cite{Kesseli2017} use 7 M8 dwarfs to calculate their spectral indices, compared to a sample of 184 M5 stars - so it is perhaps not surprising that the empirical spectroscopic analysis is in disagreement with all other wave bands. As such we conclude that SCR 1746 is of spectral type at least M5 but note that a later spectral-type may be possible. 


The \textit{TESS} monitoring of SCR 1746 (Figures \ref{fig:TESS} and \ref{fig:PhaseFold}) shows clear optical flares, along with a photometric modulation of relative amplitude $0.52 \pm 0.01$ at a period of $0.2292 \pm 0.0025$ days. We suggest that the photometric period observed is the rotational period of the star (as in e.g. \citealp{Gizis2017}) and can check this by comparing the inferred tangential velocity to that calculated by the broadening of spectral lines in the SALT data (Figure \ref{fig:spectra}, inset). For example, \cite{Lane2007} reconcile the photometric period of a radio-detected M dwarf with its spectroscopically inferred rotational velocity, invoking magnetically induced spots as the cause of periodic variability. Spectral lines are expected to be broadened due to rotation and from their width we can estimate $v$sin$(i)$, not accounting for the star's inclination $i$ with respect to our line of sight. We assume a simple relationship for rotational velocity to be $v$sin$(i) = \Delta\lambda c/\lambda = \mathrm{FWHM_H}_{\alpha}c/2\lambda_{\mathrm{H}\alpha} $ where the factor of 2 accounts for the effect of both blue- and red-shifted radial motion. Using the SALT data $v$sin$(i)~=~33.0~\pm~1.0$~km~s$^{-1}$
, in excellent agreement with the photometric inferred velocity of $32.2~\pm~0.9$~km~s$^{-1}$. This therefore agrees with the idea that the modulation is rotational in nature, is consistent with previous findings that fast rotators are magnetically active \citep[e.g.][]{Newton2017} and even implies a high inclination of the star with respect to our line of sight. Additionally, the flares in the \textit{TESS} data tend to occur when the source is brighter or increasing in brightness (Figure \ref{fig:PhaseFold}), also in agreement with the light curve modulation being caused by large magnetically active regions rotating into view.

The optical broad-band behaviour of SCR 1746 according to ASAS-SN (Figure \ref{fig:optLCs}) appears to show clear and regular optical flares, which again is to be expected from a mid-late M dwarf \citep{Paudel2018}. During the times of radio observations, no flares in either waveband are simultaneous to within 0.4 days of observations in the other. This demonstrates the value of simultaneous radio-optical observations to further constrain source properties, something now provided by the MeerKAT-MeerLICHT coupling. If the black hole XRB H1743$-$322 enters another outburst during ThunderKAT's lifetime, then this will provide a good opportunity for multiwavelength, simultaneous follow up. 

 The M dwarf population in general is expected to harbour many potential exoplanets, due to the abundance of hosts \citep{Henry2006} and estimated occurrence rate of $\sim$1.2 planets per star, increasing at later spectral types \citep{Hardegree-Ullman2019}. Therefore, host star variability is a crucial consideration in understanding planet habitability. Namely, the flares produced by M dwarfs can be orders of magnitude larger than solar flares \citep{Lacy1976} and, along with associated space weather events such as coronal mass ejections (CMEs), are likely to erode the atmospheres of planets \citep{Lammer2007}, affecting UV surface dosage, greenhouse warming efficiency and surface water retention \citep{Airapetian2017}. This is thought to most affect planets close enough to their host to sustain liquid water, as at this distance from an M dwarf they are expected to be tidally locked to their host, resulting in small magnetic moments and a weak magnetic shield to protect against CME plasma \citep{Khodachenko2007}. Flares and superflares have been recorded from known planet or planet-candidate hosting M dwarf systems \citep{Gunther2020,Lin2021} including the 7 planets hosted by TRAPPIST-1, a nearby (12 pc) active M8 dwarf \citep{Gillon2017, Luger2017}. Flaring has even been seen from our planet hosting, stellar nearest-neighbour Proxima Centauri \citep{Anglada-Escude2016,Damasso2020}, wherein ASKAP's unique capabilities provided identification of a solar type IV burst, whose occurrence is strongly associated with space weather events \citep{Zic2020}. Given the nature of the source, multiwavelength studies of SCR 1746 will be helpful in understanding the space weather around M dwarfs, determining the habitability of any potential orbiting planets.

\section{Conclusions}
\label{sec:Conc}

We have reported on the serendipitous detection of MKT J174641.0$-$321404, a radio flaring transient source discovered in commensal searches of images from the H1743$-$322 radio field taken by MeerKAT. This is the second galactic serendipitous transient found with MeerKAT, demonstrating the unique capabilities of the current generation of radio telescopes for commensal science.  MKT J174641.0$-$321404 is coincident with M dwarf SCR 1746$-$3214, which, based on archival optical and X-ray photometry and dedicated SALT spectroscopy, is of mid-late M spectral type, near the transition between mid M dwarfs and ultracool systems. \textit{TESS} observations of SCR 1746$-$3214 show clear stellar flares, demonstrating it is an active star, whilst the observed periodic variability is rotational in nature and likely caused by the presence of magnetically active regions. These observations emphasise the necessity for multiwavelength association in understanding the nature of the transient sky, one of the core drivers of the MeerLICHT project. 

\section*{Acknowledgements}

The authors would like to thank Suzanne Aigrain, Baptiste Klein and Sophia Vaughan for their helpful discussions.
A. Andersson, L. Rhodes and N. Eisner acknowledge the support given by the Science and Technology Facilities Council through STFC studentships. 
CJL acknowledges support from the Alfred P. Sloan foundation.
LND acknowledges support from the European Research Council (ERC) under the European Union's Horizon 2020 research and innovation programme (grant agreement No 694745).
PJG acknowledges support from NRF SARChI Grant 111692.
MeerKAT is operated by the South African Radio Astronomy Observatory (SARAO), which is a facility of the National Research Foundation, an agency of the Department of Science and Innovation. 
The MeerLICHT consortium is a partnership between Radboud University, the University of Cape Town, the Netherlands Organisation for Scientific Research (NWO), the South African Astronomical Observatory (SAAO), the University of Oxford, the University of Manchester and the University of Amsterdam, in association with and, partly supported by, the South African Radio Astronomy Observatory (SARAO), the European Research Council and the Netherlands Research School for Astronomy (NOVA).

We thank the SARAO and SAAO staff involved in obtaining the MeerKAT and SALT observations.
We acknowledge the use of the Inter-University Institute for Data Intensive Astronomy (IDIA) data intensive research cloud for data processing. IDIA is a South African university partnership involving the University of Cape Town, the University of Pretoria and the University of the Western Cape.

This research has made use of the SIMBAD database, operated at CDS, Strasbourg, France \citep{2000A&AS..143....9W}.
This work has made use of data from the European Space Agency (ESA) mission {\it Gaia} (\url{https://www.cosmos.esa.int/gaia}), processed by the {\it Gaia} Data Processing and Analysis Consortium (DPAC, \url{https://www.cosmos.esa.int/web/gaia/dpac/consortium}). Funding for the DPAC has been provided by national institutions, in particular the institutions participating in the {\it Gaia} Multilateral Agreement \citep{Collaboration2016,Collaboration2018,Bailer-Jones2018}. 

This research has made use of Astropy, an Astronomy-based, community-developed Python package \citep{astropy:2013,astropy:2018}
\section*{Data Availability}

All data required to reproduce results, figures and calculations herein can be found at the relevant GitHub repository: \url{https://github.com/AnderssonAstro/MKT-J174641.0-321404-Paper-Figures} 
Please cite the following Digital Object Identifier if any of the code and data are useful to your research: \url{https://doi.org/10.5281/zenodo.6346829}



\bibliographystyle{mnras}
\bibliography{DraftBib.bib} 

\begin{thebibliography}{}
\makeatletter
\relax
\def\mn@urlcharsother{\let\do\@makeother \do\$\do\&\do\#\do\^\do\_\do\%\do\~}
\def\mn@doi{\begingroup\mn@urlcharsother \@ifnextchar [ {\mn@doi@}
  {\mn@doi@[]}}
\def\mn@doi@[#1]#2{\def\@tempa{#1}\ifx\@tempa\@empty \href
  {http://dx.doi.org/#2} {doi:#2}\else \href {http://dx.doi.org/#2} {#1}\fi
  \endgroup}
\def\mn@eprint#1#2{\mn@eprint@#1:#2::\@nil}
\def\mn@eprint@arXiv#1{\href {http://arxiv.org/abs/#1} {{\tt arXiv:#1}}}
\def\mn@eprint@dblp#1{\href {http://dblp.uni-trier.de/rec/bibtex/#1.xml}
  {dblp:#1}}
\def\mn@eprint@#1:#2:#3:#4\@nil{\def\@tempa {#1}\def\@tempb {#2}\def\@tempc
  {#3}\ifx \@tempc \@empty \let \@tempc \@tempb \let \@tempb \@tempa \fi \ifx
  \@tempb \@empty \def\@tempb {arXiv}\fi \@ifundefined
  {mn@eprint@\@tempb}{\@tempb:\@tempc}{\expandafter \expandafter \csname
  mn@eprint@\@tempb\endcsname \expandafter{\@tempc}}}

\bibitem[\protect\citeauthoryear{Airapetian, Glocer, Khazanov, Loyd, France,
  Sojka, Danchi  \& Liemohn}{Airapetian et~al.}{2017}]{Airapetian2017}
Airapetian V.~S.,  Glocer A.,  Khazanov G.~V.,  Loyd R. O.~P.,  France K.,
  Sojka J.,  Danchi W.~C.,   Liemohn M.~W.,  2017, \mn@doi [The Astrophysical
  Journal] {10.3847/2041-8213/836/1/l3}, 836, L3

\bibitem[\protect\citeauthoryear{Anglada-Escud{\'{e}}
  et~al.,}{Anglada-Escud{\'{e}} et~al.}{2016}]{Anglada-Escude2016}
Anglada-Escud{\'{e}} G.,  et~al., 2016, \mn@doi [Nature] {10.1038/nature19106},
  536, 437

\bibitem[\protect\citeauthoryear{{Astropy Collaboration} et~al.,}{{Astropy
  Collaboration} et~al.}{2013}]{astropy:2013}
{Astropy Collaboration} et~al., 2013, \mn@doi [\aap]
  {10.1051/0004-6361/201322068}, \href
  {http://adsabs.harvard.edu/abs/2013A%26A...558A..33A} {558, A33}

\bibitem[\protect\citeauthoryear{{Astropy Collaboration} et~al.,}{{Astropy
  Collaboration} et~al.}{2018}]{astropy:2018}
{Astropy Collaboration} et~al., 2018, \mn@doi [\aj] {10.3847/1538-3881/aabc4f},
  \href {https://ui.adsabs.harvard.edu/abs/2018AJ....156..123A} {156, 123}

\bibitem[\protect\citeauthoryear{Bailer-Jones, Rybizki, Fouesneau, Mantelet  \&
  Andrae}{Bailer-Jones et~al.}{2018}]{Bailer-Jones2018}
Bailer-Jones C. A.~L.,  Rybizki J.,  Fouesneau M.,  Mantelet G.,   Andrae R.,
  2018, \mn@doi [The Astronomical Journal] {10.3847/1538-3881/aacb21}, 156, 58

\bibitem[\protect\citeauthoryear{Bannister, Murphy, Gaensler, Hunstead  \&
  Chatterjee}{Bannister et~al.}{2011}]{Bannister2011}
Bannister K.~W.,  Murphy T.,  Gaensler B.~M.,  Hunstead R.~W.,   Chatterjee S.,
   2011, \mn@doi [Monthly Notices of the Royal Astronomical Society]
  {10.1111/j.1365-2966.2010.17938.x}, 412, 634

\bibitem[\protect\citeauthoryear{Bastian}{Bastian}{1990}]{Bastian1990}
Bastian T.~S.,  1990, {Radio emission from flare stars},
  \mn@doi{10.1007/BF00156794}

\bibitem[\protect\citeauthoryear{Benz \& Guedel}{Benz \&
  Guedel}{1994}]{Benz1994}
Benz A.,  Guedel M.,  1994, Astronomy and Astrophysics, 285, 621

\bibitem[\protect\citeauthoryear{Benz, Alef  \& Guedel}{Benz
  et~al.}{1995}]{Benz1995}
Benz A.,  Alef W.,   Guedel M.,  1995, Astronomy and Astrophysics, 298, 187

\bibitem[\protect\citeauthoryear{Berger}{Berger}{2002}]{Berger2002}
Berger E.,  2002, \mn@doi [The Astrophysical Journal] {10.1086/340301}, 572,
  503

\bibitem[\protect\citeauthoryear{Berger et~al.,}{Berger
  et~al.}{2001}]{Berger2001}
Berger E.,  et~al., 2001, \mn@doi [Nature] {10.1038/35066514}, 410, 338

\bibitem[\protect\citeauthoryear{Berger et~al.,}{Berger
  et~al.}{2010}]{Berger2010}
Berger E.,  et~al., 2010, \mn@doi [Astrophysical Journal]
  {10.1088/0004-637X/709/1/332}, 709, 332

\bibitem[\protect\citeauthoryear{Bhandari et~al.,}{Bhandari
  et~al.}{2018}]{Bhandari2018}
Bhandari S.,  et~al., 2018, \mn@doi [Monthly Notices of the Royal Astronomical
  Society] {10.1093/mnras/sty1157}, 478, 1784

\bibitem[\protect\citeauthoryear{Bloemen et~al.,}{Bloemen
  et~al.}{2016}]{Bloemen2016}
Bloemen S.,  et~al., 2016, in Ground-based and Airborne Telescopes VI.
  International Society for Optics and Photonics, p. 990664,
  \mn@doi{10.1117/12.2232522}, \url
  {https://www.spiedigitallibrary.org/conference-proceedings-of-spie/9906/990664/MeerLICHT-and-BlackGEM--custom-built-telescopes-to-detect-faint/10.1117/12.2232522.full}

\bibitem[\protect\citeauthoryear{Bower, Saul, Bloom, Bolatto, Filippenko, Foley
   \& Perley}{Bower et~al.}{2007}]{Bower2007}
Bower G.~C.,  Saul D.,  Bloom J.~S.,  Bolatto A.,  Filippenko A.~V.,  Foley
  R.~J.,   Perley D.,  2007, \mn@doi [The Astrophysical Journal]
  {10.1086/519831}, 666, 346

\bibitem[\protect\citeauthoryear{Boyd, Winters, Henry, Jao, Finch, Subasavage
  \& Hambly}{Boyd et~al.}{2011}]{Boyd2011}
Boyd M.~R.,  Winters J.~G.,  Henry T.~J.,  Jao W.~C.,  Finch C.~T.,  Subasavage
  J.~P.,   Hambly N.~C.,  2011, \mn@doi [Astronomical Journal]
  {10.1088/0004-6256/142/1/10}, 142, 10

\bibitem[\protect\citeauthoryear{Bramall et~al.,}{Bramall
  et~al.}{2012}]{Bramall2012}
Bramall D.~G.,  et~al., 2012, in Ground-based and Airborne Instrumentation for
  Astronomy IV. International Society for Optics and Photonics, p. 84460A,
  \mn@doi{10.1117/12.925935}, \url
  {https://www.spiedigitallibrary.org/conference-proceedings-of-spie/8446/84460A/The-SALT-HRS-spectrograph--instrument-integration-and-laboratory-test/10.1117/12.925935.full}

\bibitem[\protect\citeauthoryear{Buckley, Swart  \& Meiring}{Buckley
  et~al.}{2006}]{Buckley2006}
Buckley D. A.~H.,  Swart G.~P.,   Meiring J.~G.,  2006, in Ground-based and
  Airborne Telescopes. International Society for Optics and Photonics, p.
  62670Z, \mn@doi{10.1117/12.673750}, \url
  {https://www.spiedigitallibrary.org/conference-proceedings-of-spie/6267/62670Z/Completion-and-commissioning-of-the-Southern-African-Large-Telescope/10.1117/12.673750.full}

\bibitem[\protect\citeauthoryear{Burgasser \& Putman}{Burgasser \&
  Putman}{2005}]{Burgasser2005}
Burgasser A.~J.,  Putman M.~E.,  2005, \mn@doi [The Astrophysical Journal]
  {10.1086/429788}, 626, 486

\bibitem[\protect\citeauthoryear{{Callingham} et~al.,}{{Callingham}
  et~al.}{2021}]{2021NatAs.tmp..196C}
{Callingham} J.~R.,  et~al., 2021, \mn@doi [Nature Astronomy]
  {10.1038/s41550-021-01483-0}, \href
  {https://ui.adsabs.harvard.edu/abs/2021NatAs.tmp..196C} {}

\bibitem[\protect\citeauthoryear{Camilo et~al.,}{Camilo
  et~al.}{2018}]{Camilo2018}
Camilo F.,  et~al., 2018, \mn@doi [ApJ] {10.3847/1538-4357/AAB35A}, 856, 180

\bibitem[\protect\citeauthoryear{Collaboration et~al.,}{Collaboration
  et~al.}{2016}]{Collaboration2016}
Collaboration G.,  et~al., 2016, \mn@doi [Astronomy and Astrophysics]
  {10.1051/0004-6361/201629272}, 595, A1

\bibitem[\protect\citeauthoryear{Collaboration et~al.,}{Collaboration
  et~al.}{2018}]{Collaboration2018}
Collaboration G.,  et~al., 2018, \mn@doi [Astronomy and Astrophysics]
  {10.1051/0004-6361/201833051}, 616, A1

\bibitem[\protect\citeauthoryear{Cram \& Giampapa}{Cram \&
  Giampapa}{1987}]{Cram1987}
Cram L.~E.,  Giampapa M.~S.,  1987, \mn@doi [The Astrophysical Journal]
  {10.1086/165829}, 323, 316

\bibitem[\protect\citeauthoryear{Cram \& Mullan}{Cram \&
  Mullan}{1979}]{Cram1979}
Cram L.~E.,  Mullan D.~J.,  1979, \mn@doi [The Astrophysical Journal]
  {10.1086/157532}, 234, 579

\bibitem[\protect\citeauthoryear{Crause et~al.,}{Crause
  et~al.}{2014}]{Crause2014}
Crause L.~A.,  et~al., 2014, in Ground-based and Airborne Instrumentation for
  Astronomy V. International Society for Optics and Photonics, p. 91476T,
  \mn@doi{10.1117/12.2055635}, \url
  {https://www.spiedigitallibrary.org/conference-proceedings-of-spie/9147/91476T/Performance-of-the-Southern-African-Large-Telescope-SALT-High-Resolution/10.1117/12.2055635.full}

\bibitem[\protect\citeauthoryear{Crawford et~al.,}{Crawford
  et~al.}{2016}]{Crawford2016}
Crawford S.~M.,  et~al., 2016, in Ground-based and Airborne Instrumentation for
  Astronomy VI. International Society for Optics and Photonics, p. 99082L,
  \mn@doi{10.1117/12.2232653}, \url
  {https://www.spiedigitallibrary.org/conference-proceedings-of-spie/9908/99082L/Data-reductions-and-data-quality-for-the-high-resolution-spectrograph/10.1117/12.2232653.full}

\bibitem[\protect\citeauthoryear{Cutri et~al.,}{Cutri et~al.}{2003}]{Cutri2003}
Cutri R.~M.,  et~al., 2003, VizieR On-line Data Catalog: II/246. Originally
  published in: 2003yCat.2246....0C, p. II/246

\bibitem[\protect\citeauthoryear{Cutri et~al.,}{Cutri et~al.}{2013}]{Cutri2013}
Cutri R.~M.,  et~al., 2013, VizieR On-line Data Catalog: II/328. Originally
  published in: IPAC/Caltech (2013), p. II/328

\bibitem[\protect\citeauthoryear{D'Elia et~al.,}{D'Elia
  et~al.}{2013}]{DElia2013}
D'Elia V.,  et~al., 2013, \mn@doi [Astronomy and Astrophysics]
  {10.1051/0004-6361/201220863}, 551, 142

\bibitem[\protect\citeauthoryear{Damasso et~al.,}{Damasso
  et~al.}{2020}]{Damasso2020}
Damasso M.,  et~al., 2020, \mn@doi [Science Advances] {10.1126/sciadv.aax7467},
  6, eaax7467

\bibitem[\protect\citeauthoryear{Dawson et~al.,}{Dawson
  et~al.}{2013}]{Dawson2013}
Dawson K.~S.,  et~al., 2013, \mn@doi [The Astronomical Journal]
  {10.1088/0004-6256/145/1/10}, 145, 10

\bibitem[\protect\citeauthoryear{Douglas et~al.,}{Douglas
  et~al.}{2014}]{Douglas2014}
Douglas S.~T.,  et~al., 2014, \mn@doi [Astrophysical Journal]
  {10.1088/0004-637X/795/2/161}, 795, 161

\bibitem[\protect\citeauthoryear{Driessen et~al.,}{Driessen
  et~al.}{2020}]{Driessen2020}
Driessen L.~N.,  et~al., 2020, \mn@doi [Monthly Notices of the Royal
  Astronomical Society] {10.1093/mnras/stz3027}, 491, 560

\bibitem[\protect\citeauthoryear{Driessen, Williams, McDonald, Stappers,
  Buckley, Fender  \& Woudt}{Driessen et~al.}{2021}]{Driessen2021}
Driessen L.~N.,  Williams D. R.~A.,  McDonald I.,  Stappers B.~W.,  Buckley D.
  A.~H.,  Fender R.~P.,   Woudt P.~A.,  2021, \mn@doi [MNRAS]
  {10.5281/zenodo.4515114}, 000, 1

\bibitem[\protect\citeauthoryear{{Driessen} et~al.,}{{Driessen}
  et~al.}{2022}]{2022MNRAS.tmp..746D}
{Driessen} L.~N.,  et~al., 2022, \mn@doi [\mnras] {10.1093/mnras/stac756},
  \href {https://ui.adsabs.harvard.edu/abs/2022MNRAS.tmp..746D} {}

\bibitem[\protect\citeauthoryear{Dulk}{Dulk}{1985}]{Dulk1985}
Dulk G.~A.,  1985, \mn@doi [Annual Review of Astronomy and Astrophysics]
  {10.1146/annurev.aa.23.090185.001125}, 23, 169

\bibitem[\protect\citeauthoryear{Evans et~al.,}{Evans et~al.}{2014}]{Evans2014}
Evans P.~A.,  et~al., 2014, \mn@doi [Astrophysical Journal, Supplement Series]
  {10.1088/0067-0049/210/1/8}, 210, 8

\bibitem[\protect\citeauthoryear{Evans et~al.,}{Evans et~al.}{2020}]{Evans2020}
Evans P.~A.,  et~al., 2020, \mn@doi [The Astrophysical Journal Supplement
  Series] {10.3847/1538-4365/ab7db9}, 247, 54

\bibitem[\protect\citeauthoryear{Fender \& Bell}{Fender \&
  Bell}{2011}]{Fender2011}
Fender R.~P.,  Bell M.~E.,  2011, Bulletin of the Astronomical Society of
  India, 39, 315

\bibitem[\protect\citeauthoryear{Fender et~al.,}{Fender
  et~al.}{2016}]{Fender2016}
Fender R.,  et~al., 2016, in Proceedings of Science.  (\mn@eprint {arXiv}
  {1711.04132}), \mn@doi{10.22323/1.277.0013}, \url {http://pos.sissa.it/}

\bibitem[\protect\citeauthoryear{Filippazzo, Rice, Faherty, Cruz, {Van Gordon}
  \& Looper}{Filippazzo et~al.}{2015}]{Filippazzo2015}
Filippazzo J.~C.,  Rice E.~L.,  Faherty J.,  Cruz K.~L.,  {Van Gordon} M.~M.,
  Looper D.~L.,  2015, \mn@doi [Astrophysical Journal]
  {10.1088/0004-637X/810/2/158}, 810, 158

\bibitem[\protect\citeauthoryear{Gillon et~al.,}{Gillon
  et~al.}{2017}]{Gillon2017}
Gillon M.,  et~al., 2017, \mn@doi [Nature] {10.1038/nature21360}, 542, 456

\bibitem[\protect\citeauthoryear{Gizis, Paudel, Mullan, Schmidt, Burgasser  \&
  Williams}{Gizis et~al.}{2017}]{Gizis2017}
Gizis J.~E.,  Paudel R.~R.,  Mullan D.,  Schmidt S.~J.,  Burgasser A.~J.,
  Williams P. K.~G.,  2017, \mn@doi [The Astrophysical Journal]
  {10.3847/1538-4357/aa7da0}, 845, 33

\bibitem[\protect\citeauthoryear{G{\"{u}}del}{G{\"{u}}del}{2002}]{Gudel2002}
G{\"{u}}del M.,  2002, \mn@doi [Annual Review of Astronomy and Astrophysics]
  {10.1146/annurev.astro.40.060401.093806}, 40, 217

\bibitem[\protect\citeauthoryear{Guedel \& Benz}{Guedel \&
  Benz}{1993}]{Guedel1993}
Guedel M.,  Benz A.~O.,  1993, \mn@doi [The Astrophysical Journal]
  {10.1086/186766}, 405, L63

\bibitem[\protect\citeauthoryear{G{\"{u}}nther et~al.,}{G{\"{u}}nther
  et~al.}{2020}]{Gunther2020}
G{\"{u}}nther M.~N.,  et~al., 2020, \mn@doi [The Astronomical Journal]
  {10.3847/1538-3881/ab5d3a}, 159, 60

\bibitem[\protect\citeauthoryear{Hardegree-Ullman, Cushing, Muirhead  \&
  Christiansen}{Hardegree-Ullman et~al.}{2019}]{Hardegree-Ullman2019}
Hardegree-Ullman K.~K.,  Cushing M.~C.,  Muirhead P.~S.,   Christiansen J.~L.,
  2019, \mn@doi [The Astronomical Journal] {10.3847/1538-3881/ab21d2}, 158, 75

\bibitem[\protect\citeauthoryear{Henry}{Henry}{2006}]{Henry2006}
Henry T.~J.,  2006, in Proceedings of the International Astronomical Union.
  p.~299, \mn@doi{10.1017/S174392130700419X}, \url
  {https://ui.adsabs.harvard.edu/abs/2007IAUS..240..299H/abstract}

\bibitem[\protect\citeauthoryear{Hyman, Lazio, Kassim, Ray, Markwardt  \&
  Yusef-Zadeh}{Hyman et~al.}{2005}]{Hyman2005}
Hyman S.~D.,  Lazio T. J.~W.,  Kassim N.~E.,  Ray P.~S.,  Markwardt C.~B.,
  Yusef-Zadeh F.,  2005, \mn@doi [Nature] {10.1038/nature03400}, 434, 50

\bibitem[\protect\citeauthoryear{Jaeger, Hyman, Kassim  \& Lazio}{Jaeger
  et~al.}{2012}]{Jaeger2012}
Jaeger T.~R.,  Hyman S.~D.,  Kassim N.~E.,   Lazio T.~J.,  2012, \mn@doi
  [Astronomical Journal] {10.1088/0004-6256/143/4/96}, 143, 96

\bibitem[\protect\citeauthoryear{{Jenkins} et~al.,}{{Jenkins}
  et~al.}{2016}]{jenkinsSPOC2016}
{Jenkins} J.~M.,  et~al., 2016, in Software and Cyberinfrastructure for
  Astronomy IV. p. 99133E, \mn@doi{10.1117/12.2233418}

\bibitem[\protect\citeauthoryear{Johnston et~al.,}{Johnston
  et~al.}{2007}]{Johnston2007}
Johnston S.,  et~al., 2007, \mn@doi [Publications of the Astronomical Society
  of Australia] {10.1071/AS07033}, 24, 174

\bibitem[\protect\citeauthoryear{Kesseli, West, Veyette, Harrison, Feldman  \&
  Bochanski}{Kesseli et~al.}{2017}]{Kesseli2017}
Kesseli A.~Y.,  West A.~A.,  Veyette M.,  Harrison B.,  Feldman D.,   Bochanski
  J.~J.,  2017, \mn@doi [The Astrophysical Journal Supplement Series]
  {10.3847/1538-4365/aa656d}, 230, 16

\bibitem[\protect\citeauthoryear{Khodachenko et~al.,}{Khodachenko
  et~al.}{2007}]{Khodachenko2007}
Khodachenko M.~L.,  et~al., 2007, \mn@doi [Astrobiology]
  {10.1089/ast.2006.0127}, 7, 167

\bibitem[\protect\citeauthoryear{Kirkpatrick, Henry  \& {McCarthy, Donald
  W.}}{Kirkpatrick et~al.}{1991}]{Kirkpatrick1991}
Kirkpatrick J.~D.,  Henry T.~J.,   {McCarthy, Donald W.} J.,  1991, \mn@doi
  [The Astrophysical Journal Supplement Series] {10.1086/191611}, 77, 417

\bibitem[\protect\citeauthoryear{Kochanek et~al.,}{Kochanek
  et~al.}{2017}]{Kochanek2017}
Kochanek C.~S.,  et~al., 2017, \mn@doi [Publications of the Astronomical
  Society of the Pacific] {10.1088/1538-3873/aa80d9}, 129, 104502

\bibitem[\protect\citeauthoryear{Lacy, Moffett  \& Evans}{Lacy
  et~al.}{1976}]{Lacy1976}
Lacy C.~H.,  Moffett T.~J.,   Evans D.~S.,  1976, \mn@doi [The Astrophysical
  Journal Supplement Series] {10.1086/190358}, 30, 85

\bibitem[\protect\citeauthoryear{Lacy, Chandler, Kimball, Myers, Nyland, Witz
  \& Team}{Lacy et~al.}{2019}]{Lacy2019}
Lacy M.,  Chandler C.,  Kimball A.,  Myers S.,  Nyland K.,  Witz S.,   Team V.,
   2019, Journal of Astronomical Instrumentation, 523

\bibitem[\protect\citeauthoryear{Lammer et~al.,}{Lammer
  et~al.}{2007}]{Lammer2007}
Lammer H.,  et~al., 2007, \mn@doi [Astrobiology] {10.1089/ast.2006.0128}, 7,
  185

\bibitem[\protect\citeauthoryear{Lane et~al.,}{Lane et~al.}{2007}]{Lane2007}
Lane C.,  et~al., 2007, \mn@doi [The Astrophysical Journal] {10.1086/523041},
  668, L163

\bibitem[\protect\citeauthoryear{Levinson, Ofek, Waxman  \& Gal‐Yam}{Levinson
  et~al.}{2002}]{Levinson2002}
Levinson A.,  Ofek E.~O.,  Waxman E.,   Gal‐Yam A.,  2002, \mn@doi [The
  Astrophysical Journal] {10.1086/341866}, 576, 923

\bibitem[\protect\citeauthoryear{Lin et~al.,}{Lin et~al.}{2021}]{Lin2021}
Lin C.-L.,  et~al., 2021, \mn@doi [The Astronomical Journal]
  {10.3847/1538-3881/abf933}, 162, 11

\bibitem[\protect\citeauthoryear{{Lomb}}{{Lomb}}{1976}]{Lomb76}
{Lomb} N.~R.,  1976, \mn@doi [\apss] {10.1007/BF00648343}, \href
  {https://ui.adsabs.harvard.edu/abs/1976Ap&SS..39..447L} {39, 447}

\bibitem[\protect\citeauthoryear{Lovell}{Lovell}{1969}]{Lovell1969}
Lovell B.,  1969, \mn@doi [Nature] {10.1038/2221126a0}, 222, 1126

\bibitem[\protect\citeauthoryear{Lovell, Whipple  \& Solomon}{Lovell
  et~al.}{1963}]{Lovell1963}
Lovell B.,  Whipple F.~L.,   Solomon L.~H.,  1963, \mn@doi [Nature]
  {10.1038/198228a0}, 198, 228

\bibitem[\protect\citeauthoryear{Luger et~al.,}{Luger et~al.}{2017}]{Luger2017}
Luger R.,  et~al., 2017, \mn@doi [Nature Astronomy] {10.1038/s41550-017-0129},
  1, 0129

\bibitem[\protect\citeauthoryear{Mcmullin, Waters, Schiebel, Young  \&
  Golap}{Mcmullin et~al.}{2007}]{Mcmullin2007}
Mcmullin J.~P.,  Waters B.,  Schiebel D.,  Young W.,   Golap K.,  2007, ASPC,
  376, 127

\bibitem[\protect\citeauthoryear{Mohan \& Rafferty}{Mohan \&
  Rafferty}{2015}]{Mohan2015}
Mohan N.,  Rafferty D.,  2015, Astrophysics Source Code Library, p.
  ascl:1502.007

\bibitem[\protect\citeauthoryear{Monet et~al.,}{Monet et~al.}{2003}]{Monet2003}
Monet D.~G.,  et~al., 2003, The Astronomical Journal, 125, 984

\bibitem[\protect\citeauthoryear{Mooley et~al.,}{Mooley
  et~al.}{2016}]{Mooley2016}
Mooley K.~P.,  et~al., 2016, \mn@doi [The Astrophysical Journal]
  {10.3847/0004-637x/818/2/105}, 818, 105

\bibitem[\protect\citeauthoryear{Murphy et~al.,}{Murphy
  et~al.}{2013}]{Murphy2013}
Murphy T.,  et~al., 2013, \mn@doi [Publications of the Astronomical Society of
  Australia] {10.1017/pasa.2012.006}, 30

\bibitem[\protect\citeauthoryear{Murphy et~al.,}{Murphy
  et~al.}{2017}]{Murphy2017}
Murphy T.,  et~al., 2017, \mn@doi [Monthly Notices of the Royal Astronomical
  Society] {10.1093/mnras/stw3087}, 466, 1944

\bibitem[\protect\citeauthoryear{Newton, Irwin, Charbonneau, Berlind, Calkins
  \& Mink}{Newton et~al.}{2017}]{Newton2017}
Newton E.~R.,  Irwin J.,  Charbonneau D.,  Berlind P.,  Calkins M.~L.,   Mink
  J.,  2017, \mn@doi [The Astrophysical Journal] {10.3847/1538-4357/834/1/85},
  834, 85

\bibitem[\protect\citeauthoryear{Ofek, Frail, Breslauer, Kulkarni, Chandra,
  Gal-Yam, Kasliwal  \& Gehrels}{Ofek et~al.}{2011}]{Ofek2011}
Ofek E.~O.,  Frail D.~A.,  Breslauer B.,  Kulkarni S.~R.,  Chandra P.,  Gal-Yam
  A.,  Kasliwal M.~M.,   Gehrels N.,  2011, \mn@doi [Astrophysical Journal]
  {10.1088/0004-637X/740/2/65}, 740, 65

\bibitem[\protect\citeauthoryear{Offringa \& Smirnov}{Offringa \&
  Smirnov}{2017}]{Offringa2017}
Offringa A.~R.,  Smirnov O.,  2017, \mn@doi [Monthly Notices of the Royal
  Astronomical Society] {10.1093/mnras/stx1547}, 471, 301

\bibitem[\protect\citeauthoryear{Offringa, {Van De Gronde}  \&
  Roerdink}{Offringa et~al.}{2012}]{Offringa2012}
Offringa A.~R.,  {Van De Gronde} J.~J.,   Roerdink J.~B.,  2012, \mn@doi
  [Astronomy and Astrophysics] {10.1051/0004-6361/201118497}, 539, A95

\bibitem[\protect\citeauthoryear{Offringa et~al.,}{Offringa
  et~al.}{2014}]{Offringa2014}
Offringa A.~R.,  et~al., 2014, \mn@doi [Monthly Notices of the Royal
  Astronomical Society] {10.1093/mnras/stu1368}, 444, 606

\bibitem[\protect\citeauthoryear{Osten}{Osten}{2007}]{Osten2007}
Osten R.~A.,  2007, in Proceedings of Science. Sissa Medialab Srl (\mn@eprint
  {arXiv} {0801.2573}), \url {http://arxiv.org/abs/0801.2573}

\bibitem[\protect\citeauthoryear{Page et~al.,}{Page et~al.}{2012}]{Page2012}
Page M.~J.,  et~al., 2012, \mn@doi [Monthly Notices of the Royal Astronomical
  Society] {10.1111/j.1365-2966.2012.21706.x}, 426, 903

\bibitem[\protect\citeauthoryear{Paudel, Gizis, Mullan, Schmidt, Burgasser,
  Williams  \& Berger}{Paudel et~al.}{2018}]{Paudel2018}
Paudel R.~R.,  Gizis J.~E.,  Mullan D.~J.,  Schmidt S.~J.,  Burgasser A.~J.,
  Williams P. K.~G.,   Berger E.,  2018, \mn@doi [The Astrophysical Journal]
  {10.3847/1538-4357/aab8fe}, 858, 55

\bibitem[\protect\citeauthoryear{Pietka, Fender  \& Keane}{Pietka
  et~al.}{2015}]{Pietka2015}
Pietka M.,  Fender R.~P.,   Keane E.~F.,  2015, \mn@doi [Monthly Notices of the
  Royal Astronomical Society] {10.1093/mnras/stu2335}, 446, 3687

\bibitem[\protect\citeauthoryear{Prasad et~al.,}{Prasad
  et~al.}{2016}]{Prasad2016}
Prasad P.,  et~al., 2016, {The AARTFAAC All-Sky Monitor: System Design and
  Implementation} (\mn@eprint {arXiv} {1609.04205}),
  \mn@doi{10.1142/S2251171716410087}, \url {https://arxiv.org/abs/1609.04205v1}

\bibitem[\protect\citeauthoryear{Pritchard et~al.,}{Pritchard
  et~al.}{2021}]{Pritchard2021}
Pritchard J.,  et~al., 2021, \mn@doi [Monthly Notices of the Royal Astronomical
  Society] {10.1093/mnras/stab299}, 502, 5438

\bibitem[\protect\citeauthoryear{Radcliffe, Beswick, Thomson, Garrett, Barthel
  \& Muxlow}{Radcliffe et~al.}{2019}]{Radcliffe2019}
Radcliffe J.~F.,  Beswick R.~J.,  Thomson A.~P.,  Garrett M.~A.,  Barthel
  P.~D.,   Muxlow T.~W.,  2019, \mn@doi [Monthly Notices of the Royal
  Astronomical Society] {10.1093/mnras/stz2748}, 490, 4024

\bibitem[\protect\citeauthoryear{Rajpurohit, Reyl{\'{e}}, Allard, Homeier,
  Schultheis, Bessell  \& Robin}{Rajpurohit et~al.}{2013}]{Rajpurohit2013}
Rajpurohit A.~S.,  Reyl{\'{e}} C.,  Allard F.,  Homeier D.,  Schultheis M.,
  Bessell M.~S.,   Robin A.~C.,  2013, Astronomy and Astrophysics, 556

\bibitem[\protect\citeauthoryear{{Ricker} et~al.,}{{Ricker}
  et~al.}{2015}]{ricker15}
{Ricker} G.~R.,  et~al., 2015, \mn@doi [Journal of Astronomical Telescopes,
  Instruments, and Systems] {10.1117/1.JATIS.1.1.014003}, \href
  {https://ui.adsabs.harvard.edu/abs/2015JATIS...1a4003R} {1, 014003}

\bibitem[\protect\citeauthoryear{Roulston, Green  \& Kesseli}{Roulston
  et~al.}{2020}]{Roulston2020}
Roulston B.~R.,  Green P.~J.,   Kesseli A.~Y.,  2020, \mn@doi [The
  Astrophysical Journal Supplement Series] {10.3847/1538-4365/aba1e7}, 249, 34

\bibitem[\protect\citeauthoryear{Rowlinson et~al.,}{Rowlinson
  et~al.}{2019}]{Rowlinson2019}
Rowlinson A.,  et~al., 2019, \mn@doi [Astronomy and Computing]
  {10.1016/j.ascom.2019.03.003}, 27, 111

\bibitem[\protect\citeauthoryear{Sarbadhicary et~al.,}{Sarbadhicary
  et~al.}{2020}]{Sarbadhicary2020}
Sarbadhicary S.~K.,  et~al., 2020, arXiv: 2009.05056

\bibitem[\protect\citeauthoryear{{Scargle}}{{Scargle}}{1982}]{Scargle82}
{Scargle} J.~D.,  1982, \mn@doi [\apj] {10.1086/160554}, \href
  {https://ui.adsabs.harvard.edu/abs/1982ApJ...263..835S} {263, 835}

\bibitem[\protect\citeauthoryear{Sebastian et~al.,}{Sebastian
  et~al.}{2021}]{Sebastian2021}
Sebastian D.,  et~al., 2021, \mn@doi [Astronomy and Astrophysics]
  {10.1051/0004-6361/202038827}, 645, 100

\bibitem[\protect\citeauthoryear{Shappee et~al.,}{Shappee
  et~al.}{2014}]{Shappee2014}
Shappee B.~J.,  et~al., 2014, \mn@doi [Astrophysical Journal]
  {10.1088/0004-637X/788/1/48}, 788, 48

\bibitem[\protect\citeauthoryear{Skrutskie et~al.,}{Skrutskie
  et~al.}{2006}]{Skrutskie2006}
Skrutskie M.~F.,  et~al., 2006, \mn@doi [The Astronomical Journal]
  {10.1086/498708}, 131, 1163

\bibitem[\protect\citeauthoryear{Slee, Willes  \& Robinson}{Slee
  et~al.}{2003}]{Slee2003}
Slee O.~B.,  Willes A.~J.,   Robinson R.~D.,  2003, \mn@doi [Publications of
  the Astronomical Society of Australia] {10.1071/AS03011}, 20, 257

\bibitem[\protect\citeauthoryear{{Spitzer Science Center}}{{Spitzer Science
  Center}}{2009}]{SpitzerScienceCenter2009}
{Spitzer Science Center} 2009, Vizier, p. II/293

\bibitem[\protect\citeauthoryear{Stassun et~al.,}{Stassun
  et~al.}{2019}]{Stassun2019}
Stassun K.~G.,  et~al., 2019, \mn@doi [The Astronomical Journal]
  {10.3847/1538-3881/ab3467}, 158, 138

\bibitem[\protect\citeauthoryear{Stewart et~al.,}{Stewart
  et~al.}{2016}]{Stewart2016}
Stewart A.~J.,  et~al., 2016, \mn@doi [Monthly Notices of the Royal
  Astronomical Society] {10.1093/mnras/stv2797}, 456, 2321

\bibitem[\protect\citeauthoryear{Swinbank et~al.,}{Swinbank
  et~al.}{2015}]{Swinbank2015}
Swinbank J.~D.,  et~al., 2015, \mn@doi [Astronomy and Computing]
  {10.1016/j.ascom.2015.03.002}, 11, 25

\bibitem[\protect\citeauthoryear{Tingay et~al.,}{Tingay
  et~al.}{2012}]{Tingay2012}
Tingay S.~J.,  et~al., 2012, in Proceedings of Science. Sissa Medialab Srl
  (\mn@eprint {arXiv} {1212.1327}), \mn@doi{10.22323/1.163.0036}

\bibitem[\protect\citeauthoryear{{Van Haarlem} et~al.,}{{Van Haarlem}
  et~al.}{2013}]{VanHaarlem2013}
{Van Haarlem} M.~P.,  et~al., 2013, \mn@doi [Astronomy and Astrophysics]
  {10.1051/0004-6361/201220873}, 556, A2

\bibitem[\protect\citeauthoryear{Varghese, Obenberger, Dowell  \&
  Taylor}{Varghese et~al.}{2019}]{Varghese2019}
Varghese S.~S.,  Obenberger K.~S.,  Dowell J.,   Taylor G.~B.,  2019, \mn@doi
  [The Astrophysical Journal] {10.3847/1538-4357/ab07c6}, 874, 151

\bibitem[\protect\citeauthoryear{Villadsen \& Hallinan}{Villadsen \&
  Hallinan}{2019}]{Villadsen2019}
Villadsen J.,  Hallinan G.,  2019, \mn@doi [The Astrophysical Journal]
  {10.3847/1538-4357/aaf88e}, 871, 214

\bibitem[\protect\citeauthoryear{Walkowicz, Hawley  \& West}{Walkowicz
  et~al.}{2004}]{Walkowicz2004}
Walkowicz L.~M.,  Hawley S.~L.,   West A.~A.,  2004, \mn@doi [Publications of
  the Astronomical Society of the Pacific] {10.1086/426792}, 116, 1105

\bibitem[\protect\citeauthoryear{{Wang} et~al.,}{{Wang}
  et~al.}{2021}]{2021ApJ...920...45W}
{Wang} Z.,  et~al., 2021, \mn@doi [\apj] {10.3847/1538-4357/ac2360}, \href
  {https://ui.adsabs.harvard.edu/abs/2021ApJ...920...45W} {920, 45}

\bibitem[\protect\citeauthoryear{{Wenger} et~al.,}{{Wenger}
  et~al.}{2000}]{2000A&AS..143....9W}
{Wenger} M.,  et~al., 2000, \mn@doi [\aaps] {10.1051/aas:2000332}, \href
  {https://ui.adsabs.harvard.edu/abs/2000A&AS..143....9W} {143, 9}

\bibitem[\protect\citeauthoryear{{West}, {Walkowicz}  \& {Hawley}}{{West}
  et~al.}{2005}]{2005PASP..117..706W}
{West} A.~A.,  {Walkowicz} L.~M.,   {Hawley} S.~L.,  2005, \mn@doi [\pasp]
  {10.1086/431368}, \href
  {https://ui.adsabs.harvard.edu/abs/2005PASP..117..706W} {117, 706}

\bibitem[\protect\citeauthoryear{Williams et~al.,}{Williams
  et~al.}{2020}]{Williams2020}
Williams D.~R.,  et~al., 2020, \mn@doi [Monthly Notices of the Royal
  Astronomical Society: Letters] {10.1093/mnrasl/slz152}, 491, L29

\bibitem[\protect\citeauthoryear{Zic et~al.,}{Zic et~al.}{2019}]{Zic2019}
Zic A.,  et~al., 2019, \mn@doi [Monthly Notices of the Royal Astronomical
  Society] {10.1093/mnras/stz1684}, 488, 559

\bibitem[\protect\citeauthoryear{Zic et~al.,}{Zic et~al.}{2020}]{Zic2020}
Zic A.,  et~al., 2020, \mn@doi [The Astrophysical Journal]
  {10.3847/1538-4357/abca90}, 905, 23

\makeatother
\end{thebibliography}








\bsp	
\label{lastpage}
\end{document}